\documentclass[11pt]{article}
\pdfoutput=1
\usepackage{cite}
\usepackage{scalerel}
\usepackage{amsmath}
\usepackage{amsfonts}

\usepackage{mathabx}
\usepackage{extarrows}

\def\hybrid{
        \topmargin -20pt
        \oddsidemargin 0pt
        \headheight 0pt \headsep 0pt
        \textwidth 6.25in 
        \textheight 9.5in 
        \marginparwidth .875in
        \parskip 5pt plus 1pt \jot = 1.5ex}

\hybrid

 \csname
@addtoreset\endcsname{equation}{section}

\def\cL{{\cal L}}
\def\cD{\mathfrak{D}}
\def\cF{{\cal F}}
\def\cA{{\cal A}}
\def\cM{{\cal M}}
\def\cH{{\cal H}}

\def\del{\partial}

\def\be{\begin{equation}}
\def\ee{\end{equation}}

\begin{document}

\begin{titlepage}
\rightline{}
\rightline{April 2019}
\rightline{HU-EP-19/08}
\begin{center}
\vskip 1.5cm
 {\Large \bf{ Leibniz Gauge Theories and Infinity Structures}}
\vskip 1.7cm

{\large\bf {Roberto Bonezzi and Olaf Hohm}}
\vskip 1.6cm

{\it  Institute for Physics, Humboldt University Berlin,\\
 Zum Gro\ss en Windkanal 6, D-12489 Berlin, Germany}\\
 ohohm@physik.hu-berlin.de\\
roberto.bonezzi@physik.hu-berlin.de 
\vskip .1cm

\vskip .2cm

\end{center}

\bigskip\bigskip
\begin{center} 
\textbf{Abstract}

\end{center} 
\begin{quote}

We formulate gauge theories based on Leibniz(-Loday) algebras 
and uncover their underlying mathematical structure. 
Various special cases have been developed in the context of 
gauged supergravity and exceptional field theory. These are based on 
`tensor hierarchies', which  describe towers of $p$-form gauge fields transforming under 
non-abelian gauge symmetries and which have been constructed 
up to low levels. Here we define `infinity-enhanced Leibniz algebras' 
that guarantee the existence of consistent tensor hierarchies to arbitrary level. 
We contrast these algebras with strongly homotopy Lie algebras ($L_{\infty}$ algebras), 
which can be used to define topological field theories for which all curvatures vanish. 
Any infinity-enhanced Leibniz algebra carries an associated $L_{\infty}$ algebra, which 
we discuss.

\end{quote} 
\vfill
\setcounter{footnote}{0}
\end{titlepage}

\tableofcontents

\newpage 

\section{Introduction}

In this paper we construct the general gauge theory of 
Leibniz-Loday algebras \cite{LODAY,Baraglia,Strobl1,Lavau:2017tvi,Hohm:2018ybo,Kotov:2018vcz}, 
which are algebraic structures generalizing the notion of Lie algebras. These structures  have appeared 
in the context of duality covariant formulations of gauged supergravity \cite{deWit:2008ta,deWit:2002vt,deWit:2004nw,deWit:2005hv} 
and of string/M-theory 
\cite{Hull:2009zb,Coimbra:2011ky,Berman:2012vc,Cederwall:2013oaa,Hohm:2013nja,Hohm:2013pua,Hohm:2013vpa,Hohm:2013uia,Hohm:2014fxa,Abzalov:2015ega,Musaev:2015ces,Hohm:2015xna,Berman:2015rcc,Lee:2014mla}. 
Such gauge theories and their associated tensor hierarchies (towers of $p$-form 
gauge fields transforming under non-abelian gauge symmetries) have so far been constructed on a case-by-case basis up to the level 
needed in a given number of dimensions. 
Our goal  is to develop gauge theories based on Leibniz(-Loday) algebras in all generality and to axiomatize 
the underlying mathematical structure that guarantees consistency of the tensor hierarchies up to arbitrary levels. 
This gauge theory construction has a certain degree of universality in that it is 
based on an algebraic structure 
encoding  the most general bilinear `product' defining  transformations whose closure  is governed 
by the same product, thereby generalizing the adjoint action of a Lie algebra.

We begin by discussing this notion of universality as a way of introducing Leibniz algebras.  
There is a definite sense in which the most general algebraic structures defining (infinitesimal) symmetries are Lie algebras; indeed, we usually take the 
notion of continuous symmetries and Lie algebras to be synonymous. 
Let us briefly recall a `proof' of this lore: suppose 
we are given infinitesimal variations $\delta_{\lambda}\phi^i$ 
that leave an action $I[\phi^i]$ invariant, i.e., 
 \be
  0 \ = \ \delta_{\lambda}I[\phi^i] \ = \ \int \frac{\delta I}{\delta \phi^i}\,\delta_{\lambda}\phi^i \;, 
 \ee
where $\phi^i$ collectively denotes all fields. We can now act with another symmetry variation and antisymmetrize, which yields 
 \be
 \begin{split}
  0 \ = \ (\delta_{\lambda_1}\delta_{\lambda_2}-\delta_{\lambda_2}\delta_{\lambda_1})I[\phi^i] \ &= \ 
  \int \Big(2\,\frac{\delta^2I}{\delta \phi^i\delta \phi^j}\,\delta_{\lambda_{[1}}\phi^i \,\delta_{\lambda_{2]}}\phi^j 
  + \frac{\delta I}{\delta \phi^i}\,[\delta_{\lambda_1},\delta_{\lambda_2}]\phi^i\Big)
  \;. 
 \end{split}
 \ee
Since the second variational derivative is symmetric, the first term vanishes and  
we infer 
 \be
  0\ = \ 
  \int  \frac{\delta I}{\delta \phi^i}\,[\delta_{\lambda_1},\delta_{\lambda_2}]\phi^i \;. 
 \ee
But this means that $[\delta_{\lambda_1},\delta_{\lambda_2}]\phi^i$ 
is also an invariance. Therefore, symmetries `close', so that we can write 
 \be\label{closure}
  [\delta_{\lambda_1},\delta_{\lambda_2}]\phi^i \ = \ \delta_{[\lambda_1,\lambda_2]}\phi^i\;, 
 \ee
where we now take the $\lambda$ to parameterize all invariances of $I[\phi^i]$ and $[\cdot, \cdot]$ on the right-hand side to be defined by this relation. 
Since the left-hand side of (\ref{closure}) is just a commutator, 
the Jacobi identity $[[\delta_{\lambda_1},\delta_{\lambda_2}],\delta_{\lambda_3}]+{\rm cycl.}=0$ is identically satisifed. 
It follows that the antisymmetric bracket $[\cdot, \cdot]$ on the right-hand side of (\ref{closure}) also satisfies the Jacobi identity and hence 
defines a Lie algebra. 

The above proof has several loopholes. For instance, the bracket (\ref{closure}) could be field-dependent or closure could hold only 
`on-shell', i.e., modulo trivial equations-of-motion symmetries. There is a well-developed machinery in (quantum) field theory 
to deal with such issues, the BV formalism \cite{Henneaux:1992ig} 
(which in turn is related to $L_{\infty}$ algebras that in this paper will play a role in a slightly different context). 
Here, however, we are concerned with another, more algebraic  loophole: the existence of `trivial symmetry parameters' whose action on fields vanishes, so that 
the Jacobi identity does not need to hold exactly for the bracket $[\cdot, \cdot]$, as long as its `Jacobiator' lies in the 
space of trivial parameters. 
We want to ask: what is the most general bilinear algebraic operation (product) defined on a vector space that gives rise to consistent 
symmetry variations that close according to the same product? 
We will now argue that such algebraic structures are Leibniz algebras: they are defined by a bilinear operation $\circ$, 
satisfying the  identity 
 \be\label{LeibnizIDIntro}
  x\circ (y\circ z) - y\circ (x\circ z) \ = \ (x\circ y)\circ z \;. 
 \ee
Given such a bilinear  operation, we can define variations 
 \be\label{Introvart}
  \delta_{x}y \ \equiv \ {\cal L}_{x}y \ \equiv \ x\circ y\;, 
 \ee
where we introduced the notation ${\cal L}_x$ to be used below. 
Now, the Leibniz relation (\ref{LeibnizIDIntro}) is nothing more or less than the requirement that these are consistent symmetry transformations whose 
closure is governed by $\circ$: 
 \be\label{gaugealgebra0}
 \begin{split}
  [{\cal L}_x, {\cal L}_y]z \ & \equiv \  {\cal L}_x({\cal L}_y z)- {\cal L}_y({\cal L}_x z) \\
  \ &= \ x\circ (y\circ z) -y\circ (x\circ z) \\
   \ &= \ (x\circ y)\circ z \\
    \ &= \ {\cal L}_{x\circ y}z\;. 
 \end{split}
 \ee 
In this sense, Leibniz algebras are the answer to our question above.   
In particular, we do not need to assume that the product $\circ$ is antisymmetric. If the product is antisymmetric, the Leibniz relation (\ref{LeibnizIDIntro}) coincides 
with the Jacobi identity, and hence a Lie algebra is a special case of a Leibniz algebra. 
If the product is not antisymmetric, it carries a non-vanishing symmetric part denoted by $\{\cdot,\cdot\}$.  
Symmetrizing (\ref{gaugealgebra0}) in $x, y$ we infer ${\cal L}_{\{x,y\}}z=0$ for any $z$, which implies that there is a space of trivial parameters, 
in which the symmetric part takes values. We will see that the antisymmetric part denoted by $[\cdot, \cdot]$ does not satisfy the Jacobi identity, 
but its Jacobiator yields a trivial parameter.

The reader may wonder what the significance of Leibniz algebras is, given that the symmetric part $\{\cdot,\cdot\}$, which encodes the 
deviation from a Lie algebra, acts trivially. Indeed, we will see that the space of trivial parameters forms an ideal of the antisymmetric bracket, 
hence we could pass to the quotient algebra by modding out the trivial parameters, for which the resulting bracket is antisymmetric and does 
satisfy the Jacobi identity. In this sense, it is indeed sufficient to work with Lie algebras. So why should we bother with Leibniz algebras? 
The reason is the same as for gauge symmetries in general. Gauge invariances encode redundancies of the formulation, 
and hence in principle can be disposed of by working on the `space of gauge invariant functions' or, alternatively, by `fixing a gauge'. 
But the fact of the matter is that a redundant formulation is often greatly beneficial. Typically, a gauge theory formulation is necessary 
in order to render Lorentz invariance and locality manifest. Similarly, the Leibniz algebras arising in gauged supergravity and 
exceptional field theory are necessary in order to render duality symmetries manifest. 
The price to pay is then a yet higher level of redundancy, in which one not only has equivalences between certain 
field configurations but also equivalences between equivalences, etc., leading to the notion of `higher gauge theories'. 
(See \cite{Jurco:2019woz} for a recent introduction to higher gauge theories.)

The higher gauge theory structures manifest themselves in the form of `tensor hierarchies', which arise when one attempts to mimic 
the construction of Yang-Mills theory by introducing one-form gauge fields taking values in the Leibniz algebra. 
Since the associated antisymmetric bracket $[\cdot, \cdot]$ does not obey the Jacobi identity, however, one cannot define a 
gauge covariant field strength as in Yang-Mills theory. This can be resolved by introducing two-form potentials taking values 
in the space of trivial parameters and coupling it to the naive field strength. This, in turn, requires the introduction of three-form potentials 
in order to define a covariant field strength for the two-forms, indicating a pattern that potentially continues indefinitely. 
Thus, the seemingly minor relaxation of Lie algebra structures given by Leibniz algebras has profound consequence for the 
associated gauge theories, leading to a rich structure of higher-form symmetries. In the case of exceptional field theory, 
this gives a rationale for the presence of higher-form gauge fields in M-theory. 

A core feature of the tensor hierarchy construction is that 
familiar relations from Yang-Mills theory, as closure of gauge transformations for the one-form connection or covariance of its 
naive field strength, only hold `up to higher-form gauge transformations'. The resulting structure closely resembles that of 
strongly homotopy Lie algebras ($L_{\infty}$ algebras), in which the standard Lie algebra relations only hold `up to homotopy', i.e., up to higher brackets that in turn satisfy higher Jacobi identities \cite{Zwiebach:1992ie,Lada:1992wc,Lada,Hohm:2017pnh}. 
Indeed, the relation with $L_{\infty}$ algebras has already been elaborated in a number of publications, 
see \cite{Palmer:2013pka,Lavau:2014iva,Saemann:2017rjm,Ritter:2015ymv,Cagnacci:2018buk}. 
More generally, in the mathematics literature it is well established that many algebraic structures or operations have 
`infinity' versions, in which the standard relations only hold `up to homotopy'. (See, for instance, \cite{Fialowski} and \cite{Vallette} for a pedagogical introduction.) 
Our goal here is to identify the `infinity structure'  
that underlies tensor hierarchies. 
While  $L_{\infty}$ algebras and tensor hierarchies are  closely related, it turns out that by themselves the former do not 
provide a proper axiomatization of the latter. Further structures are needed, beyond the graded antisymmetric brackets of 
$L_{\infty}$ algebras, in order to define the most general tensor hierarchies. A first step towards the mathematical characterization of such structures 
was taken by Strobl, who introduced `enhanced Leibniz algebras' \cite{Strobl:2016aph,Strobl:2019hha}, which extend a Leibniz algebra 
by an additional vector space, together with a new algebraic operation satisfying suitable compatibility conditions with the Leibniz product. 
This structure is sufficient in order to define tensor hierarchies that end with two-forms. 
Here we go beyond this by identifying the mathematical structure that can be used to define tensor hierarchies up to arbitrary degrees, 
which we term `infinity enhanced Leibniz algebras'. In this we rely heavily on the results obtained in \cite{Hohm:2015xna,Wang:2015hca}.

In the following we briefly display some of our core technical results. As a first step, the vector space of the Leibniz algebra 
is extended to a chain complex $X=\bigoplus_{n=0}^{\infty}X_n$ with a degree-$(-1)$ differential $\mathfrak{D}$ (satisfying $\mathfrak{D}^2=0$), 
of which the Leibniz algebra forms the degree-zero subspace $X_0$. While the Leibniz product is only defined on $X_0$, we postulate a degree-$(+1)$  
graded symmetric map $\bullet :X\otimes X\rightarrow X$, satisfying suitable (compatibility-)conditions with the differential $\mathfrak{D}$ 
and the Leibniz product, as for instance
 $\{x,y\}  =   \tfrac{1}{2}\mathfrak{D}(x\bullet y)$ for  $x,y\in X_0$.  Defining the operator $\iota_x u:= x\bullet u$ for $x\in X_0$ and $u$ 
 arbitrary one can then define a Lie derivative 
 with respect to $\lambda\in X_0$ in analogy to Cartan's `magic formula' for the Lie derivative acting on differential forms, 
  \be
   {\cal L}_{\lambda} \ \equiv \ \iota_{\lambda}\,\mathfrak{D} \ + \ \mathfrak{D}\,\iota_{\lambda}\;. 
  \ee
This Lie derivative  satisfies all familiar relations as a consequence of the general axioms we formulate. 
One can then define a gauge theory for a set of $p$-form gauge fields of arbitrary rank, each form taking values in $X_{p-1}$. 
The one-form gauge field taking values in $X_0$ plays a distinguished role. The resulting formulas can be 
written very efficiently in terms of formal sums for the remaining gauge fields $\cA:=\sum_{p=2}^\infty A_p$, 
the curvatures $\cF:=\sum_{p=2}^\infty\cF_p$ and the 
Chern-Simons-type forms $\mathit{\Omega}:=\sum_{p=2}^\infty\Omega_p$, defined by  
\begin{equation}
\Omega_n(A)=\tfrac{(-1)^n}{(n-1)!}\,(\iota_A)^{n-2}\big[dA-\tfrac{1}{n}\,A\circ A\big]    \;. 
\end{equation}
We will show that the curvatures defined  by 
  \be
   \cF\ = \ \sum_{N=0}^\infty\frac{(-\iota_{\cA})^N}{(N+1)!}\Big[(D+\mathfrak{D})\cA+  (N+1)\mathit{\Omega}\Big]\;, 
 \ee  
where $D=d-{\cal L}_{A_1}$ is the covariant derivative, satisfy the Bianchi identity 
\begin{equation}
\begin{split}
D\cF \ + \ \tfrac12\,\cF\bullet\cF \ = \ \mathfrak{D}\cF  \;.
\end{split}    
\end{equation}
Writing it out in terms of differential forms, the Bianchi identity takes a hierarchical form that 
relates the covariant exterior derivative of the $p$-form field strength ${\cal F}_p$ to 
the differential $\mathfrak{D}$ of the $(p+1)$-form field strength ${\cal F}_{p+1}$, c.f.~(\ref{Bianchirepeated}) below. 
The gauge covariance of the lowest field strength ${\cal F}_2$  
then implies by induction gauge covariance of all field strengths.

The rest of this paper is organized as follows. 
In sec.~2 we discuss general results on Leibniz algebras to set the stage for the construction of gauge theories. 
In order to keep the paper self-contained and accessible we then present a step-by-step construction of the associated tensor hierarchy up 
to some low form-degree. In sec.~3 we use various observations made along the way, generalized further to arbitrary degrees, 
in order to motivate the general axioms of `infinity-enhanced Leibniz algebras' that will be used in sec.~4 to construct 
exact tensor hierarchies. These are not restricted to finite degrees, and we prove the consistency of the tensor hierarchy to all orders. 
This construction is then contrasted in sec.~5 with topological field theories based on $L_{\infty}$ algebras, which are consistent, without the 
need to introduce infinity-enhanced Leibniz algebras, by virtue of all field strengths being zero. 
We conclude in sec.~6 with a brief summary and outlook, while the Appendices include some technical details 
needed for the proof of the Bianchi identity, as well as a discussion of $L_{\infty}$ algebras associated to Leibniz algebras.

\emph{Note added in proof:}\\
During the review stage of this  article there have been further developments in the understanding of infinity-enhanced Leibniz algebras \cite{Lavau:2019oja,Bonezzi:2019bek}. 
In particular, this gives an improved motivation for the axioms of infinity-enhanced Leibniz algebras as being obtained through a derived construction from 
a differential graded Lie algebra and a subsequent truncation to spaces of non-negative degree. 
This is reflected in the new subsection 3.2 of this paper.

\section{Generalities on Leibniz gauge theories}

In this section we develop Leibniz algebras and discuss the first few steps needed in order 
to define their associated gauge theories. Specifically, this requires an extension of the original vector space 
on which the Leibniz algebra is defined by a `space of trivial parameters' together with a new algebraic operation. 
Eventually, this construction will be extended to a graded sum of vector spaces with a differential (chain complex) 
and a bilinear graded symmetric operation. The results of this section will motivate the general axioms to be 
presented in the next section.

\subsection{Leibniz algebras}

As outlined  in the introduction, a Leibniz (or Loday) algebra is a vector space $V$ equipped with a `product' or 2-bracket $\circ$ 
satisfying for $x,y,z\in V$ the Leibniz identity (\ref{LeibnizIDIntro}), which we here rewrite as 
 \be\label{LeibnizID}
  x\circ (y\circ z) \ = \ (x\circ y)\circ z + y\circ (x\circ z)\;. 
 \ee
This form makes it clear that the symmetry variations defined by (\ref{Introvart}), i.e., $\delta_{x}y =  {\cal L}_{x}y  =  x\circ y$, 
act according to the Leibniz rule on the product $\circ$, hence explaining the name `Leibniz algebra'. 
(Sometimes this is referred to as `left Leibniz algebra'. One could also introduce a `right Leibniz algebra', where a vector acts 
from the right.)
Similarly, it follows that the product is covariant under these transformations: 
\be\label{covarianceprop}
\begin{split}
 \delta_{x}(y\circ z) \ &\equiv \ \delta_xy\circ z +y\circ \delta_xz \\
 \ &= \  (x \circ y)\circ z +y\circ (x\circ z) \\
 \ &= \ x\circ (y\circ z) \\
  \ &= \ {\cal L}_x(y\circ z)\;. 
\end{split}
\ee 
Conversely, demanding that the product $\circ$ defines a symmetry operation 
that is covariant with respect to itself uniquely leads to the notion of a Leibniz algebra. 

We can now derive some further consequences from the Leibniz relations, 
in particular from the closure relation (\ref{gaugealgebra0}),  
 \be\label{gaugealgebra}
 \begin{split}
  [{\cal L}_x, {\cal L}_y]z \ = \ {\cal L}_{x\circ y}z\;. 
 \end{split}
 \ee 
Defining 
 \be
 \begin{split}
  \{x,y\} \ &\equiv \ \tfrac{1}{2}(x\circ y + y\circ x)\;, \\
  [x,y] \  &\equiv  \ \tfrac{1}{2}(x\circ y - y\circ x)\;, 
 \end{split} 
 \ee 
and symmetrizing (\ref{gaugealgebra}) in $x, y$ we have 
 \be
  [{\cal L}_x,{\cal L}_y]z  
  \  = \ {\cal L}_{[x,y]}z\;, 
 \ee
and   
 \be\label{symmistrivial}
  {\cal L}_{\{x,y\}}z \ = \ 0 \quad \forall x,y \;. 
 \ee
Thus, the antisymmetric part defines the `structure constants' of the more conventional (antisymmetric) gauge algebra, 
but we will see shortly that it does not satisfy the Jacobi identity. 
Indeed, as discussed in the introduction, we infer from (\ref{symmistrivial}) that 
in general there is a notion of `trivial gauge parameters', given by the symmetric part, 
so that it is sufficient that the  `Jacobiator' is trivial in this sense. 
We can now prove that the `Jacobiator' of the bracket $[\,,]$ is trivial in that 
 \be
  {\rm Jac}(x_1, x_2, x_3) \ \equiv \ 3[[x_{[1}, x^{}_2],x_{3]}] \ = \ \{x_{[1}\circ x^{}_{2},  x_{3]}\}\;. 
 \ee
For the proof we suppress the total antisymmetrization in $1,2,3$. We then need to establish: 
 \be
  6[x_1\circ x_2, x_3] - 2\{x_1\circ x_2, x_3\} \ = \ 0\;, 
 \ee
where we multiplied by $2$ for convenience.  
This relation is verified by writing out the brackets, using total antisymmetry and the Leibniz identity (\ref{LeibnizID}) 
in the last step: 
 \be
  \begin{split}
  6[x_1\circ x_2, x_3] - 2\{x_1\circ x_2, x_3\}  = 
   & \ 3\,(x_1\circ x_2)\circ x_3 - 3\,x_3\circ (x_1\circ x_2) \\
   &\ -(x_1\circ x_2)\circ x_3 - x_3\circ (x_1\circ x_2) \\
   = &\ 2\,(x_1\circ x_2)\circ x_3-4\,x_3\circ (x_1\circ x_2)\\
   = &\ 2\,(x_1\circ x_2)\circ x_3+2\, x_2\circ (x_1\circ x_3) - 2\, x_1\circ (x_2\circ x_3) \\
   = &\ 0\;.  
  \end{split}
 \ee

It should be emphasized that the above structure is only non-trivial iff the symmetric pairing 
$\{\,,\}$ takes values  in a proper subspace of $V$, for otherwise we had with (\ref{symmistrivial})
that $\forall x: {\cal L}_xz=0$, i.e., that the product is trivial. If $\{\,,\}=0$, we have a 
Lie algebra. More generally, the above structures define an $L_{\infty}$ algebra with `2-bracket' $\ell_2(x,y)=[x,y]$. 
Provided the space of trivial gauge parameters forms an ideal, 
this follows directly from Theorem 2 in \cite{Hohm:2017cey}. 
In order to prove that the trivial parameters form an ideal 
we have to show that the bracket of an arbitrary vector $z$ 
with $\{x,y\}$ is again trivial, i.e., writable in terms of  $\{\,,\}$. 
To this end, we use that the covariance property (\ref{covarianceprop}) implies the covariance 
of the symmetric pairing: 
 \be\label{covarianceofproduct}
 \begin{split}
  z\circ \{x,y\} \ = \ \{ z\circ x, y\} + \{x,z\circ y\} 
  \;. 
 \end{split} 
 \ee
Since this also equals 
  \be
    z\circ \{x,y\} \ = \ [z, \{x,y\} ]+\{z,  \{x,y\} \}\;, 
  \ee
 we have     
 \be\label{idealprop}
  [z, \{x,y\} ] \ = \ \{ z\circ x, y\} + \{x,z\circ y\} - \{z,  \{x,y\} \}\;. 
 \ee
This completes the proof that the bracket of a trivial element with an arbitrary vector $z$ is itself trivial 
and hence that the space of trivial vectors forms an ideal. Therefore, as mentioned in the introduction, 
we could pass to the quotient algebra in which one identifies two vectors that differ by a `trivial' vector, which then defines a Lie algebra. 
In applications, however, this can typically not be done in a duality covariant manner.

It will next turn out to be convenient to parameterize the space of trivial parameters more 
explicitly, so that the symmetric part of the product can be viewed as the image of a linear nilpotent 
operator of another algebraic operation. 
Specifically, we introduce a vector space $U$ and a linear operator $\mathfrak{D}:U\rightarrow V$, 
so that 
 \be\label{DREL}
  \{x,y\} \ = \  \tfrac{1}{2}\cD(x\bullet y)\;, 
 \ee
where $\bullet$ is a symmetric bilinear map $V\otimes V\rightarrow U$, 
and the factor of $\tfrac{1}{2}$ is for convenience. 
This relation is motivated by `infinity' structures such as $L_{\infty}$ algebras, 
where a nilpotent differential on a chain complex governs the homotopy versions of algebraic relations, 
and also will turn out to be necessary in order to define  tensor hierarchies explicitly. 
One can assume (\ref{DREL}) without loss of generality. For instance, 
if $\{\,,\}$ lives in a subspace of $V$, 
we can take $U$ to be isomorphic to this subspace and $\cD$ the inclusion map that views an element 
of $U$ as an element of $V$. However, $\cD$ can be more general, and in particular have a non-trivial 
kernel. In examples, $\cD$ typically emerges naturally as a non-trivial operator. 

Let us spell out some further assumptions on the space $U$ and then derive some consequences of (\ref{DREL}). First, (\ref{symmistrivial}) in combination with 
(\ref{DREL}) implies ${\cal L}_{\cD(x\bullet y)}z=\cD(x\bullet y)\circ z=0$ for all $x, y\in V$. 
We will assume that the space $U$ has been chosen so as to precisely encode the trivial parameters in that  
 \be\label{DisTrivial}
  \forall u\in U:\;\;\; \cD u\circ x \ = \ {\cal L}_{\cD u}x \ = \ 0\;. 
 \ee
Immediate corollaries are
 \be
  \forall x\in V\,u\in U:\;\; \{ x, \cD u\} \ = \ \{\cD u, x\} \ = \ \tfrac{1}{2}x\circ \cD u\;, 
 \ee 
and therefore with (\ref{DREL}) 
 \be\label{LEMMMMMAAAA}
  \forall x\in V\, u\in U:\;\; \cD (x\bullet \cD u) \ = \ x\circ \cD u\;. 
 \ee
This means that the Leibniz product of an arbitrary vector with any trivial ($\cD $ exact) vector is itself $\cD $ exact and hence trivial. 
Another consequence is derived by setting $x=\cD v$ in (\ref{LEMMMMMAAAA}), 
 \be
  \cD (\cD v\bullet  \cD u) \ = \ \cD v\circ \cD u \ = \ 0\;. 
 \ee 
Put differently, the $\bullet$ product of two $\cD $ exact elements takes values in the kernel of $\cD $: 
 \be\label{DKernel}
  \cD v \bullet  \cD u \ \in \ {\rm Ker}(\cD )\;. 
 \ee
 
In the remainder of this subsection we make the assumption that the kernel of $\cD $ is trivial in order to 
exemplify the  resulting structures in the simplest possible setting and to connect to the 
`enhanced Leibniz algebras' discussed recently in  \cite{Strobl:2019hha}. 
Put differently, we assume a structure given by the 2-term chain complex 
\be\label{SIMPLEexactsequence}
  U \; \xrightarrow{\cD } \; V\;, 
 \ee
so that $\cD u=0$ implies $u=0$.  Although somewhat degenerate, this setup already allows us to exhibit 
some features that later will recur in the general context. 
 
Our first goal is to prove that the bilinear operation $\bullet $ is covariant w.r.t.~a natural action of the Leibniz algebra 
on $u\in U$ given by   
 \be\label{Lieaction}
  {\cal L}_{x}u \ \equiv \ x\bullet \cD u \;. 
 \ee
Thus, we want to prove that 
 \be\label{INVARIANCEPair}
  \delta_{z}(x\bullet y) \ \equiv \ (z\circ x)\bullet y + x\bullet (z\circ y) \ = \ z\bullet \cD (x\bullet y) \;.
 \ee
To this end we employ the covariance of the Leibniz product 
$\circ$ w.r.t.~its own action, as expressed in (\ref{covarianceprop}), (\ref{covarianceofproduct}), to compute 
 \be
  \delta_{z}(\cD (x\bullet y)) \ = \ 2\,\delta_{z}\{x,y\} \ = \ 2\,z\circ \{x, y\} \ = \ z\circ \cD (x\bullet y) \ = \ 
  \cD (z\bullet \cD (x\bullet y))\;, 
 \ee
where we used (\ref{LEMMMMMAAAA}) in the last step. Since, by definition of variations,  the left-hand side equals 
$\cD (\delta_{z}(x\bullet y))$, we have established: 
 \be\label{projInvariance}
  \cD \big(\delta_{z}(x\bullet y)-z\bullet \cD (x\bullet y)\big) \ = \ 0\;. 
 \ee
Since we assumed the kernel of $\cD $ to be trivial, (\ref{INVARIANCEPair}) follows, as we wanted to prove.

We can now prove that the action (\ref{Lieaction}) of the Leibniz algebra on $U$ closes according to the Leibniz product. 
We first note the general fact that the following combination lives in the kernel of $\cD $: 
 \be\label{KErnelProp}
  x\bullet (y\circ \cD a) - y\bullet (x\circ \cD a) - (x\circ y)\bullet \cD a \ \in \ {\rm Ker}(\cD )\;. 
 \ee
This is verified by acting with $\cD $, using the defining relation (\ref{DREL}) and writing out 
the Leibniz products:  
 \be
 \begin{split}
  2&\big(\{ x, y\circ \cD a\} - \{y, x\circ \cD a\} - \{x\circ y, \cD a\}\big) \\
  & \ = \ 
  x\circ (y\circ \cD a) + (y\circ \cD a)\circ x - y\circ (x\circ \cD a) - (x\circ \cD a)\circ y
  -(x\circ y)\circ \cD a - \cD a\circ (x\circ y)\\
  & \ = \ 
   (y\circ \cD a)\circ x  - (x\circ \cD a)\circ y
   - \cD a\circ (x\circ y)\\
  & \ = \ 
   (y\circ \cD a)\circ x  - (x\circ \cD a)\circ y\\
  & \ = \ 0\;, 
 \end{split}
 \ee 
where we used the Leibniz algebra relations (\ref{LeibnizID}) 
and the properties of trivial parameters (\ref{DisTrivial}). 
This completes the proof of (\ref{KErnelProp}). 
Since we assume that the kernel of $\cD $ is trivial, it follows that the expression to the left of (\ref{KErnelProp}) 
vanishes. Closure of (\ref{Lieaction}) then follows: 
 \be
  \begin{split}
   [{\cal L}_{z_1}, {\cal L}_{z_2}]u \ &= \ z_1\bullet \cD (z_2\bullet \cD u)
   -z_1\bullet \cD (z_2\bullet \cD u)\\
   \ &= \ z_1\bullet (z_2\circ \cD u)-z_2\bullet (z_1\circ \cD u) \\
   \ &= \ (z_1\circ z_2)\bullet \cD u\\ 
   \ &= \ {\cal L}_{z_1\circ z_2}u\;, 
  \end{split}
 \ee 
where we used (\ref{LEMMMMMAAAA}) in the second line.

\subsection{Generalization to non-trivial kernel}

We will now relax some of the assumptions above. First, 
we allow $\cD $ to have a non-trivial kernel. 
This implies that the chain complex (\ref{SIMPLEexactsequence}) has to be extended 
by an additional space and a new differential $\cD $ whose image parameterizes the 
kernel of the previous differential. Adopting a notation for vector spaces labelled by 
their degree (w.r.t.~the grading of the chain complex to be developed shortly), 
we consider the complex 
 \be\label{generalchain}
 \begin{split}
  \cdots  \quad \rightarrow &\quad X_2
  \quad \xrightarrow{\cD _2} \quad X_1 \quad \xrightarrow{\cD _1} 
  \quad X_{0}\;\;,
 \end{split} 
 \ee 
where we inserted,  as a subscript,  the space on which $\cD $ acts. 
The graded vector space, together with the linear maps $\cD $, forms a chain complex, 
which means that $\cD ^2=0$ or, more precisely, $\cD _i\circ \cD _{i+1}=0$.\footnote{As customary, we also denote the composition
of maps by $\circ$. It should always be clear from the context whether we mean the Leibniz product or composition.}
Thus, there is a notion of \textit{homology}: the quotient space of $\cD $-closed elements modulo $\cD $-exact elements. 
In the remainder of this section we will assume this homology to be trivial, so that any $\cD $-closed element is $\cD $-exact. 
This allows us to derive relations needed for the construction of tensor hierarchies, the beginning of which will be discussed 
in the next subsection. In secs.~3 and 4 below we will then take these relations to be imposed axiomatically, so that the homology need not be trivial.

Let us now develop some relations involving elements of the new space $X_2$. 
From (\ref{DKernel}) we infer 
that the symmetric pairing of two $\cD _1$ exact elements takes values in the kernel of $\cD _1$.
Thus, by the assumption of trivial homology, the result is  $\cD $-exact --- with respect to the new $\cD _2$. 
In analogy to (\ref{DREL}) we then introduce a new bilinear operation $\bullet$ to write 
 \be\label{goodKErDEF}
  \forall a,b\in X_1\;:\quad  \cD _1 a\bullet \cD _1b \ \equiv \ -\cD _2(a\bullet \cD _1b) \ \equiv \ -\cD _2(b\bullet \cD _1a)\;, 
 \ee
where the sign is for later convenience.  Moreover, we have introduced on the 
r.h.s.~maps $\bullet\,:\; X_1\otimes X_0\rightarrow X_2$ of intrinsic degree $+1$. 
The equality of both forms on the r.h.s.~follows from the  l.h.s.~being  symmetric in $a, b$, so that 
we can assume that the r.h.s.~is also symmetric. 
Put differently, we can assume  that the antisymmetric part is $\cD $ exact and write 
 \be\label{newLeibnizRule}
  a\bullet \cD _1b \ - \  b\bullet \cD _1 a \ = \ \cD _3 (a\bullet b)\;, 
 \ee
with  $a\bullet b\in X_3$ and a new differential $\cD _3: X_3\rightarrow X_2$. 
More generally, we can anticipate the existence of a bilinear operation 
$\bullet$ of intrinsic degree 1 that is \textit{graded symmetric}, i.e., 
  \be
   \forall A, B \in X\,:\quad A\bullet B \ = \ (-1)^{|A||B|} B\bullet A\;, \qquad
   A \bullet B \ \in \ X_{|A||B|+1}\;. 
  \ee
Indeed, according to the grading for $a, b\in X_1$ the product needs to be antisymmetric, 
in agreement with the implicit definition (\ref{newLeibnizRule}).

Our next goal is to define a generalization of the Leibniz action (\ref{Lieaction}) that is valid on the entire chain complex. 
Specifically, we define a generalized Lie derivative via `Cartan's magic formula'  
 \be\label{LieGenCartan}
  {\cal L}_{z}a \ \equiv \ z\bullet \cD a + \cD (z\bullet a)\;, 
 \ee
for $z\in X_0$ and $a\in X_i$, $i>0$. The complete analogy to Cartan's formula for Lie derivatives 
of differential forms can be made manifest by introducing the map 
 \be\label{iotaAction}
   \iota_z\,:\; X_i\,\rightarrow\, X_{i+1}\;, \qquad \iota_z(a) \ \equiv \ z\bullet a\;, 
  \ee 
for $z\in X_0$, because then (\ref{LieGenCartan}) can be written as  
 \be\label{LieCARTANNN}
  {\cal L}_z \ = \ \iota_z\,\cD +\cD \,\iota_z\;. 
 \ee  
 
We will next try to establish standard relations for Lie derivatives, which in turn requires imposing 
further relations between $\cD $ and $\bullet$.  
We first show that an element of the original Leibniz algebra that is $\cD $-exact acts 
trivially according to the Cartan formula --- as it should be in view of the interpretation of $X_1$ 
as the `space of trivial parameters'. To this end we set  $z=\cD b$ and compute with (\ref{LieGenCartan}) 
 \be\label{generalTRIV}
   {\cal L}_{\cD b}a \ = \ \cD b\bullet \cD a + \cD (\cD b\bullet a) \ = \ 
   -\cD (a\bullet \cD b)+ \cD (\cD b\bullet a) \ = \ 0\;, 
 \ee
using (\ref{goodKErDEF}) and the graded commutativity of $\bullet$. (The sign choice in (\ref{goodKErDEF}) 
was made such that the trivial parameters of the action given by the Cartan formula are $\cD $-exact.)  
Another direct consequence of the definition (\ref{LieGenCartan}) is that  Lie derivatives
commute with $\cD $:   
 \be\label{DcalLCommute}
  [\cD ,{\cal L}_x] \ = \ \cD (\iota_x \cD +\cD \iota_x) 
  - (\iota_x \cD +\cD \iota_x)\cD  \ = \ 0\;. 
 \ee 
Put differently, $\cD $ is a covariant operation.

Let us now address the crucial question whether the generalized Lie derivatives (\ref{LieCARTANNN}) 
form an algebra. This can be easily seen to be the case if and only if the Lie derivative is `covariant' 
w.r.t~its own action. 
Since, as just established, $\cD $ is covariant, we expect that the Lie derivatives close 
if we assume that the operations $\bullet$ are defined so as to be covariant. 
More precisely, we demand that 
the operation (\ref{iotaAction}) 
transforms covariantly under the generalized Lie derivatives (\ref{LieGenCartan}) in the sense 
that 
 \be\label{covLieonX1}
  \delta_x(\iota_y(A)) \ \equiv \ \iota_{x\circ y}(A)+\iota_y({\cal L}_xA) \ \equiv \ 
  {\cal L}_{x}(\iota_y(A))\;. 
 \ee
The last equality is the statement of covariance. This can be rewritten as 
 \be
  {\cal L}_x\iota_y - \iota_y{\cal L}_x \ = \ \iota_{x\circ y}\;. 
 \ee  
Even shorter, and together with (\ref{DcalLCommute}), we thus have 
 \be\label{magicProperties}
  [{\cal L}_x,\cD ] \ = \ 0\;, \qquad [{\cal L}_x,\iota_y] \ = \ \iota_{x\circ y}\;. 
 \ee
This is sufficient in order to prove closure of the algebra of generalized Lie derivatives: 
 \be\label{GeneralCLOSURE}
 \begin{split}
  [{\cal L}_x, {\cal L}_y] \ &= \ {\cal L}_x(\iota_y\cD +\cD \iota_y)-(\iota_y\cD +\cD \iota_y){\cal L}_x\\
  \ &= \ ({\cal L}_x\iota_y -\iota_y{\cal L}_x)\cD  + \cD ({\cal L}_x\iota_y - \iota_y {\cal L}_x) \\
  \ &= \ \iota_{x\circ y}\cD +\cD \iota_{x\circ y}\\
  \ &= \ {\cal L}_{x\circ y}\;. 
 \end{split} 
 \ee
We recall that Cartan's formula and hence the above proof only hold 
when acting on objects in $X_i$, $i>0$. Of course, for elements in $X_0$ closure follows from the Leibniz algebra properties.

\subsection{Tensor hierarchy at low levels}\label{sec:THlowlevels}

We will now turn to the formulation of gauge theories based on algebraic structures satisfying the 
relations discussed in the previous subsection, exhibiting the first few steps in the construction of a tensor hierarchy.
In this one tries to mimic the construction of Yang-Mills 
theory: one introduces one-forms $A=A_{\mu}\,{\rm d}x^{\mu}$, with $x^{\mu}$ the coordinates of the base `spacetime' manifold, but taking values in a 
Leibniz algebra instead of a Lie algebra.
Following the standard textbook treatment of gauge theories, we next aim to define covariant derivatives 
and field strengths. 
We can define covariant derivatives for any fields in $X$ by
 \be
  D_{\mu} \ = \ \partial_{\mu} \ - \ {\cal L}_{A_{\mu}}\;, 
 \ee
with the universal form (\ref{LieGenCartan}) of the generalized Lie derivative.  
It is a quick computation using the closure relation (\ref{GeneralCLOSURE}) to verify that 
the covariant derivative transforms covariantly, i.e., according to the same Lie derivative (\ref{LieGenCartan}). 
(For an explicit display of this proof see eq.~(139) in \cite{Hohm:2019wql}.) 
Moreover, since the $\bullet$ operation is 
covariant under these  Lie derivatives we immediately have the Leibniz rule: 
 \be\label{LeibnizCovBullet}
  D_{\mu}(a\bullet b) \ = \ D_{\mu}a\bullet b \ + \  a\bullet D_{\mu}b\;, 
 \ee
for arbitrary $a, b\in X$.  

For the gauge potential $A_\mu\,$, we postulate gauge transformations 
w.r.t.~to a Leibniz-algebra valued gauge parameter $\lambda\in X_0$: 
 \be\label{LeibnizYangMills}
  \delta_{\lambda}A_{\mu} \ = \ D_{\mu}\lambda  \ \equiv \ \partial_{\mu}\lambda - A_{\mu}\circ \lambda\;. 
 \ee 
The important difference to Yang-Mills theory originates from the fact that in general these gauge transformations do not close 
by themselves. Using the Leibniz algebra relations and (\ref{DREL}) we compute for the commutator of (\ref{LeibnizYangMills}): 
 \be
 \begin{split}
 [\delta_{\lambda_1},\delta_{\lambda_2}]A_{\mu} \ = \ &\, -2\,\partial_{\mu}\lambda_{[1}\circ \lambda_{2]}
 +2\,(A_{\mu}\circ \lambda_{[1})\circ \lambda_{2]} \\
 \ =  \ &\,-\partial_{\mu}\lambda_{[1}\circ \lambda_{2]} -\lambda_{[1}\circ \partial_{\mu}\lambda_{2]}
 +2\{\lambda_{[1},\partial_{\mu}\lambda_{2]}\}\\
 &\, +A_{\mu}\circ (\lambda_{[1}\circ \lambda_{2]})-2\{\lambda_{[1},A_{\mu}\circ \lambda_{2]}\} \\
 \ = \ &\, -D_{\mu}(\lambda_{[1}\circ \lambda_{2]}) +2\{\lambda_{[1}, D_{\mu}\lambda_{2]}\} \\
 \ = \ &\, D_{\mu}[\lambda_2,\lambda_1] +\cD (\lambda_{[1} \bullet {D}_{\mu}\lambda_{2]})\;. 
 \end{split}
 \ee 
The first term on the right-hand side takes the form of $\delta_{12}A_{\mu}$, with 
$\lambda_{12}=[\lambda_2,\lambda_1]$, but the second term is inconsistent with closure. 
The notation above suggests already the resolution: one postulates a new gauge symmetry with a parameter $\lambda_{\mu}\in X_1$: 
\be\label{fullgaugetransA}
 \delta_{\lambda}A_{\mu} \ = \ D_{\mu}\lambda \ - \ \cD \lambda_{\mu}\;. 
\ee
We then have closure according to  
$[\delta_{\lambda_1},\delta_{\lambda_2}]A_{\mu} = D_{\mu}\lambda_{12}-\cD \lambda_{12\mu}$, where 
 \be
   \lambda_{12} \ = \ [\lambda_2,\lambda_1]\;, \qquad \lambda_{12\mu} \ = \ 
   \lambda_{[2} \bullet {D}_{\mu}\lambda_{1]}\;. 
 \ee

We now turn to the definition of a non-abelian field strength for $A_{\mu}$, starting 
from the ansatz
 \be
  F_{\mu\nu} \ = \ \partial_{\mu}A_{\nu}-\partial_{\nu}A_{\mu} - [A_{\mu}, A_{\nu}] \;, 
 \ee
where further (2-form) terms will be added to achieve  gauge covariance. 
The need for this modification is most efficiently shown by computing the general variation 
of the field strength and demanding covariance. Under a general variation $\delta A_{\mu}$ we compute 
 \be\label{generalvarF}
 \begin{split}
  \delta F_{\mu\nu} \ &= \ 2\,\partial_{[\mu}(\delta A_{\nu]}) -2[A_{[\mu}, \delta A_{\nu]}] \\
  \ &= \ 2(\partial_{[\mu}\delta A_{\nu]} - A_{[\mu}\circ \delta A_{\nu]}+\{A_{[\mu}, \delta A_{\nu]}\}) \\
  \ &= \ 2\, D_{[\mu}\delta A_{\nu]} + \cD (A_{[\mu} \bullet \delta A_{\nu]})\;, 
 \end{split}
 \ee 
where we used (\ref{DREL}). Thus, we have succeeded to write $\delta F_{\mu\nu}$ in 
terms of the covariant derivative of $\delta A_{\mu}$ only up to $\cD $-exact terms. 
This is now remedied by introducing a 2-form $B_{\mu\nu}\in X_1$ and completing 
the definition of the field strength as 
 \be\label{FULLF}
  \cF_{\mu\nu} \ = \ \partial_{\mu}A_{\nu}-\partial_{\nu}A_{\mu} - [A_{\mu}, A_{\nu}] +\cD B_{\mu\nu}\;. 
 \ee
It follows with (\ref{generalvarF}) that the general variation under $\delta A_{\mu}$, $\delta B_{\mu\nu}$ takes 
the form 
 \be\label{genvarF}
  \delta \cF_{\mu\nu} \ = \ 2\, D_{[\mu}\, \delta A_{\nu]} + \cD (\Delta B_{\mu\nu})\;, 
 \ee
where we defined the `covariant variations' 
 \be\label{covVARB}
  \Delta B_{\mu\nu} \ \equiv \ \delta B_{\mu\nu}+A_{[\mu} \bullet  \delta A_{\nu]}\;. 
 \ee
We will see that these covariant variations of higher forms recur in all covariant formulas.

We can now determine the gauge transformations of the 2-forms so that the field strength transforms 
covariantly. To this end we use that with (\ref{GeneralCLOSURE}) we have 
for the commutator of covariant derivatives: 
 \be\label{commCOVDER}
  [D_{\mu}, D_{\nu}] \ = \ -{\cal L}_{F_{\mu\nu}}\equiv-\cL_{\cF_{\mu\nu}}\;. 
 \ee
Note that, due to (\ref{generalTRIV}), in this formula it is immaterial whether
the field strength on the right-hand side contains the 2-form term in (\ref{FULLF}) or not. 
 Covariance of  $\cF_{\mu\nu}$ under (\ref{LeibnizYangMills}) now follows, provided 
 we postulate the following gauge transformations for $B_{\mu\nu}$, written in terms of (\ref{covVARB}), 
  \be
  \Delta_{\lambda}B_{\mu\nu} \ = \ 2\,D_{[\mu}\lambda_{\nu]} + \cF_{\mu\nu}\bullet  \lambda\;. 
 \ee
Here we also introduced a new gauge parameter  $\lambda_{\mu}\in X_1$, for which $B_{\mu\nu}$ is the gauge field. 
Indeed, with (\ref{commCOVDER}) and (\ref{genvarF}) we then compute  
 \be
 \begin{split}
  \delta_{\lambda} \cF_{\mu\nu} \ &= \ [D_{\mu}, D_{\nu}]\lambda + \cD (\Delta_{\lambda}B_{\mu\nu}) \\
  \ &= \ \lambda\circ \cF_{\mu\nu}-2\{\cF_{\mu\nu}, \lambda\}+ \cD (\Delta_{\lambda}B_{\mu\nu}) \\
  \ &= \ \lambda\circ \cF_{\mu\nu}+\cD (\Delta_{\lambda}B_{\mu\nu} - \cF_{\mu\nu} \bullet \lambda)\\
  \ &= \ {\cal L}_{\lambda} \cF_{\mu\nu}\;. 
 \end{split}
 \ee
Next, we have to prove invariance under the new shift transformation w.r.t.~$\lambda_{\mu}\in X_1$. With (\ref{fullgaugetransA}) we compute 
 \be
  \begin{split}
   \delta_{\lambda}\cF_{\mu\nu} \ &= \ -2\, D_{[\mu}\cD \lambda_{\nu]}+2\, \cD (D_{[\mu}\lambda_{\nu]})\\
   \ &= \ -2\, \partial_{[\mu}\cD \lambda_{\nu]} + 2\, A_{[\mu}\circ \cD \lambda_{\nu]} 
   +2\,\cD \big(\partial_{[\mu}\lambda_{\nu]}-A_{[\mu}\bullet \cD \lambda_{\nu]}\big) - 
   \cD (A_{[\mu}\bullet \lambda_{\nu]})  \\
   \ &= \  2\, A_{[\mu}\circ \cD \lambda_{\nu]}  -4\{ A_{[\mu}, \cD \lambda_{\nu]}\} \\
   \ &= \ -2\,  \cD \lambda_{[\nu}\circ A_{\mu]}\\
   \ &= \ 0\;, 
  \end{split}
 \ee  
using $\cD ^2=0$ and, in the last step,  (\ref{DisTrivial}).

 Having established the covariance (and invariance) properties of $\cF_{\mu\nu}$, we next ask 
 whether there is a covariant 3-form field strength for the 2-form. 
 This 3-form curvature emerges naturally upon inspecting the possible Bianchi identities  for $\cF_{\mu\nu}$. 
 In contrast to the Bianchi identity in Yang-Mills theory based on Lie algebras, the covariant curl of $\cF_{\mu\nu}$
 in general is not zero but only $\cD $-exact, thereby introducing a 3-form that is covariant (again, up to $\cD $-exact terms).  
 Specifically, we have the following generalized Bianchi identity,  
  \be\label{FIRSTBianchi}
  3\, D_{[\mu} \cF_{\nu\rho]} \ = \ \cD H_{\mu\nu\rho}\;, 
 \ee
where 
 \be\label{DEFH}
 \begin{split}
  H_{\mu\nu\rho} \ &= \ 3\Big(\partial_{[\mu} B_{\nu\rho]}-A_{[\mu}\bullet \cD  B_{\nu\rho]} 
  -\cD (A_{[\mu}\bullet B_{\nu\rho]})
  -A_{[\mu}\bullet \partial_{\nu}A_{\rho]} + \tfrac{1}{3} A_{[\mu}\bullet A_{\nu}\circ A_{\rho]}\Big) \\
  \ &= \ 3\, \big(D_{[\mu} B_{\nu\rho]} -\Omega_{\mu\nu\rho}(A)\big)\;, 
 \end{split}
 \ee 
and we introduced  the Chern-Simons three-form 
 \be
  \Omega_{\mu\nu\rho}(A) \ \equiv \ 
  A_{[\mu}\bullet \partial_{\nu}A_{\rho]} - \tfrac{1}{3} A_{[\mu}\bullet (A_{\nu}\circ A_{\rho]})\;. 
 \ee
Note that $H_{\mu\nu\rho}$ is determined by (\ref{FIRSTBianchi}) only up to contributions 
that are $\cD $-closed and hence $\cD $-exact. In (\ref{DEFH}) 
we added a $\cD $-exact term in order to build the full covariant derivative. 
Moreover, we should not expect $H_{\mu\nu\rho}$ to be fully gauge covariant, 
but only up to $\cD $-exact contributions. These, in turn, can be fixed by introducing a 3-form gauge potential. 
The proof of the Bianchi identity (\ref{FIRSTBianchi}) proceeds by a straightforwards computation, using repeatedly (\ref{LEMMMMMAAAA})
and performing similar calculations as above.\footnote{In particular, one has to use the perhaps somewhat surprising relations 
 \be\label{surprisingREL}
  3\, A_{[\mu}\circ (A_{\nu}\circ A_{\rho]}) \ = \ \cD (A_{[\mu}\bullet( A_{\nu}\circ A_{\rho]}))\;, \qquad
  A_{\mu}\circ \cD  B_{\nu\rho} \ = \ \cD (A_{\mu}\bullet \cD  B_{\nu\rho})\;. 
 \ee}

In order to further develop the general pattern of tensor hierarchies, we close this section by completing the definition of 
the 3-form curvature by introducing a 3-form potential taking values in $X_2$. To this end it is again convenient 
to inspect the general variation of $H_{\mu\nu\rho}$. First, under an arbitrary variation $\delta A_{\mu}$, 
we compute for the Chern-Simons term\footnote{Here and in the following total antisymmetrization of form indices is understood.}  
 \be
 \begin{split}
  \delta\, \Omega_{\mu\nu\rho}(A) 
  \ = \ -D_{\mu}(A_{\nu}\bullet  \delta A_{\rho})
  +\delta A_{\mu} \bullet \cF_{\nu\rho} -\delta A_{\mu}\bullet \cD B_{\nu\rho}
  -\tfrac{1}{3}\cD (A_{\mu}\bullet (A_{\nu}\bullet \delta A_{\rho}))
  \;, 
 \end{split}
 \ee
where we used the covariance relation (\ref{covLieonX1}).   
The general variation of the three-form curvature is then given by 
 \be
 \begin{split}
  \delta H_{\mu\nu\rho} \ &= \ 3\Big(D_{\mu}\delta B_{\nu\rho} -\delta \Omega_{\mu\nu\rho}
  -\delta A_{\mu} \bullet  \cD B_{\nu\rho} -\cD (\delta A_{\mu}\bullet B_{\nu\rho}) \Big) \\
   \ &= \ 3\Big(D_{\mu}\Delta B_{\nu\rho}- \delta A_{\mu} \bullet \cF_{\nu\rho} 
   -\cD \big(\delta A_{\mu}\bullet B_{\nu\rho} - \tfrac{1}{3}
   A_{\mu}\bullet (A_{\nu}\bullet \delta A_{\rho})\big)\Big)\;. 
 \end{split}
 \ee 
The first two terms on the right-hand side are covariant, but there is also a non-covariant but $\cD $-exact term.  
Again, this can be remedied by introducing $C_{\mu\nu\rho}\in X_2$ and defining 
 \be\label{FUllH}
  {\cal H}_{\mu\nu\rho} \ \equiv \ H_{\mu\nu\rho} + \cD C_{\mu\nu\rho}\;. 
 \ee 
The  general variation can then be written as 
 \be
  \delta{\cal H}_{\mu\nu\rho} \ = \  3\, D_{\mu}\Delta B_{\nu\rho}- 3\, \delta A_{\mu} \bullet \cF_{\nu\rho} 
  +\cD \Delta C_{\mu\nu\rho}\;, 
 \ee 
with the covariant variation of the 3-form 
 \be
  \Delta C_{\mu\nu\rho} \ \equiv \ \delta C_{\mu\nu\rho} - 3\,\delta A_{\mu}\bullet B_{\nu\rho} 
  +  A_{\mu}\bullet (A_{\nu}\bullet \delta A_{\rho})\;. 
 \ee 
With the relations established so far it is now a direct computation to verify 
gauge covariance of the field strength under 
  \be
  \begin{split}
   \Delta C_{\mu\nu\rho} \ &= \ 3\,D_{\mu}\Sigma_{\nu\rho} + 3\, \cF_{\mu\nu}\bullet \lambda_{\rho}
   +\lambda\bullet {\cal H}_{\mu\nu\rho} -\cD \Xi_{\mu\nu\rho} \;, \\
    \Delta B_{\mu\nu} \ &= \ 2\, D_{\mu}\lambda_{\nu} + \lambda\bullet \cF_{\mu\nu} - \cD \Sigma_{\mu\nu}\;, \\
    \delta A_{\mu} \ &= \ D_{\mu}\lambda -\cD \lambda_{\mu}\;, 
  \end{split} 
  \ee
where we introduced higher shift gauge parameters $\Sigma_{\mu\nu}\in X_2$, $\Xi_{\mu\nu\rho}\in X_3$.  
We can use this to quickly verify that there are trivial parameters of the following form  
 \be
  \begin{split}
   \lambda \ &= \ \cD \chi\;, \\
   \lambda_{\mu} \ &= \ \cD \chi_{\mu} + D_{\mu}\chi\;, \\
   \Sigma_{\mu\nu} \ &= \ \cD \chi_{\mu\nu} + 2\, D_{\mu}\chi_{\nu} - \chi\bullet \cF_{\mu\nu} \;, \\
   \Xi_{\mu\nu\rho} \ &= \ \cD  \chi_{\mu\nu\rho} + 3\,D_{\mu}\chi_{\nu\rho} 
   -3\, \cF_{\mu\nu}\bullet \chi_{\rho} + \chi\bullet {\cal H}_{\mu\nu\rho}\;. 
  \end{split}
 \ee
For this one uses (\ref{LeibnizCovBullet}) and, for the final relation,   
(\ref{newLeibnizRule}). 

It should now be fairly clear how the pattern continues: at each level (form degree) one can construct consistent gauge transformations, 
covariant curvatures, etc., that have the familiar properties up to $\cD $-exact contributions that, in turn, can be fixed by introducing forms 
of one higher degree. The exact (or closed-form) formulation of the complete tensor hierarchy will be developed in the next two sections. 

\newpage

\section{Infinity enhanced Leibniz algebra } 

In the previous section we have seen how the step-by-step construction of the tensor hierarchy proceeds in parallel to the introduction of spaces of higher degree $X_n\,$, as well as differentials $\cD $ and graded symmetric maps $\bullet$. In this section we will give a set of axioms, involving the Leibniz product and the higher structures, that define what we call an infinity enhanced Leibniz algebra. We will then show that such an algebraic structure is sufficient in order 
to construct a 
tensor hierarchy to all orders. 
Before listing the axioms, we will show how they can be motivated from the properties of the original Leibniz algebra. This will hint to a close relation with differential graded Lie algebras, that will be discussed as well.

\subsection{Motivation}

As discussed in the previous sections, the Leibniz product provides a natural notion of symmetry transformations (in the following 
often referred to as Lie derivative):
\begin{equation}
\delta_x y\equiv\cL_xy:=x\circ y\;,    
\end{equation}
that closes and is covariant, \emph{i.e.}, 
\begin{equation}
[\cL_x,\cL_y]\,z=\cL_{[x,y]}\,z\;,    \quad \cL_x(y\circ z)=(\cL_xy)\circ z+y\circ(\cL_xz)\;.
\end{equation}
The trivial action of the symmetric pairing, $\cL_{\{x,y\}}z=0$, prompts us to introduce the space $X_1$ and the first bullet operator $\bullet:X_0\otimes X_0\to X_1$
as
\begin{equation}\label{motivation-1}
x\circ y+y\circ x=\cD(x\bullet y)    \;,
\end{equation}
where the right hand side can be viewed as the definition of $\bullet\,$. Since $\cL_{\cD(x\bullet y)}\,z=0$ for any $x,y\in X_0\,$, one is led to associate triviality of the Lie derivative with $\cD$-exactness, and thus postulate
\begin{equation}\label{motivation0}
\cD u\circ x=0\;,\quad \forall u\in X_1\;,\;x\in X_0\;.    
\end{equation}
The Lie derivative
is at the core of constructing the gauge theory, in that it defines covariant derivatives and gauge variations. Since higher form gauge fields are valued in spaces $X_n$ with $n>0\,$, it is necessary to extend the definition of the Lie derivative to spaces of arbitrary degree in a way that preserves closure and covariance. In order to determine the form of $\cL_x$ acting on elements of higher degree, we notice that 
 \be
  (x\circ y)\bullet z +(x\circ z)\bullet y -x\bullet(y\circ z + z\circ y)
 \ee
is $\cD $-closed, thanks to the Leibniz property of $\circ\,$. Imposing it to be $\cD$-exact amounts to define the higher bullet $\bullet: X_0\otimes X_1\to X_2$ by 
   \be\label{motivation1}
  (x\circ y)\bullet z + (x\circ z)\bullet y -x\bullet(y\circ z + z\circ y) \ = \ \cD (x\bullet (y\bullet z))\;.
 \ee
This is motivated by the assumption that everything should be writable only in terms of $\cD $ and $\bullet\,$, and by manifest symmetry in $y\leftrightarrow z\,$.  
By defining
\begin{equation}
\cL_xa:=x\bullet\cD a+\cD(x\bullet a)\;,\quad a\in X_n\;,n>0\;,    
\end{equation}
this is equivalent to $y\bullet z$ being covariant under ${\cal L}_x\,$. Totally symmetrizing the relation \eqref{motivation1}, 
and recalling that $\bullet$ is symmetric for degree-zero objects,  we infer 
 \be
  \cD \big[x\bullet (y\bullet z)+y\bullet (z\bullet x)+z\bullet (x\bullet y)\big] \ = \ 0\;, 
 \ee
so the expression in parenthesis is $\cD$-closed, and we impose it to be $\cD$-exact. Since there is nothing writable in terms of  $\bullet$ yielding a degree $+3$ object from $(x,y,z)$, this amounts to postulating
 \be\label{motivation2}
  x\bullet (y\bullet z)+y\bullet (z\bullet x)+z\bullet (x\bullet y) \ = \ 0\;.
 \ee
By using the notation $\iota_x=x\,\bullet\,$, we notice that \eqref{motivation2} can be rewritten in the form
\begin{equation}
-\iota_x(y\bullet z)=(\iota_xy)\bullet z+y\bullet(\iota_xz) \;,   \quad x,y,z\in X_0
\end{equation}
viewed as a (twisted) Leibniz property of the operator $\iota_x\,$. This suggests the graded extension
\begin{equation}\label{motivation3}
-\iota_x (a\bullet b)=(\iota_xa)\bullet b+(-1)^{|a|}a\bullet(\iota_x b)\;,\quad x\in X_0\;,\quad a,b\in X 
\end{equation}
to the whole space. Given \eqref{motivation3}, one can act repeatedly with $\iota_{x_k}$ to prove by induction
\begin{equation}
(-1)^{|{\bf X}_n|+1}{\bf X}_n\bullet(a\bullet b)=({\bf X}_n\bullet a)\bullet b+(-1)^{|a||b|}({\bf X}_n\bullet b)\bullet a \;,\quad\forall\;a,b\in X \end{equation}
with ${\bf X}_n:=x_1\bullet(x_2\bullet(...(x_{n-1}\bullet x_n)))$ an element of degree $n-1$ generated by nested products of degree zero elements $x_i\,$, thereby suggesting the general relation
\begin{equation}\label{motivation4}
(-1)^{|a|+1}a\bullet(b\bullet c)=(a\bullet b)\bullet c+(-1)^{|b||c|}(a\bullet c)\bullet b\;.    
\end{equation}

Coming back to the properties of the $\cD$ operator, we see from \eqref{motivation0} that $\cD u\bullet\cD v$ is $\cD$-closed for $u,v\in X_1\,$, since $\cD(\cD u\bullet \cD v)=\cD u\circ \cD v+\cD v\circ\cD u\,$. Postulating this is $\cD$-exact amounts to requiring there exists a degree 2 element, depending
on $u$ and $v\,$, whose $\cD$-boundary is $\cD u\bullet\cD v\,$; we can therefore define $-u\bullet\cD v$ to be this element, so that
\begin{equation}
\cD u\bullet\cD v=-\cD(u\bullet\cD v)=-\cD(v\bullet\cD u)\;,    
\end{equation}
where the sign has been chosen such that $\cL_{\cD u}v=0\,$, maintaining triviality of the Lie derivative along $\cD$-exact elements. Antisymmetrizing the above relation one finds that $u\bullet\cD v-v\bullet\cD u$
is $\cD$-closed. Imposing again that it is $\cD$-exact allows us to define yet a higher bullet product $\bullet:X_1\otimes X_1\to X_3$ via 
\begin{equation}
u\bullet\cD v-v\bullet\cD u=\cD(u\bullet v) \;,   
\end{equation}
suggesting the twisted Leibniz property
\begin{equation}\label{motivation5}
-\cD(a\bullet b)=(\cD a)\bullet b+(-1)^{|a|}a\bullet\cD b\;,\quad |a|,|b|>0\;.    
\end{equation}
From the relations \eqref{motivation3} and \eqref{motivation5} it is possible to prove covariance of the product $a\bullet b$ under the action of the Lie derivative when neither $a$ nor $b$ have degree zero, as  will be shown explicitly later. As we have already mentioned, \eqref{motivation1} ensures covariance of $y\bullet z$ under the Lie derivative when both $y$ and $z$ have degree zero. On the other hand, when only one argument has degree zero, \eqref{motivation3} and \eqref{motivation5} only allow us to determine
\begin{equation}
\begin{split}
\cL_x(y\bullet u)+\cL_y(x\bullet u)&= x\bullet\cD(y\bullet u)+y\bullet\cD(x\bullet u)+\cD\big[x\bullet(y\bullet u)+y\bullet(x\bullet u)\big]\\
&= x\bullet\cL_yu+y\bullet\cL_xu-x\bullet(y\bullet\cD u)-y\bullet(x\bullet\cD u)-\cD\big[(x\bullet y)\bullet u\big]\\
&=x\bullet\cL_yu+y\bullet\cL_xu+\cD(x\bullet y)\bullet u\\
&= \big[(\cL_xy)\bullet u+y\bullet\cL_xu\big]+\big[(\cL_yx)\bullet u+x\bullet\cL_yu\big]\;,
\end{split}    
\end{equation}
showing that one needs to demand
\begin{equation}\label{motivation6}
\cD(x_{[1}\bullet(x_{2]}\bullet u))=2\,x_{[2}\bullet\cD(x_{1]}\bullet u)+x_{[2}\bullet(x_{1]}\bullet\cD u)+[x_1,x_2]\bullet u \;,\quad |x_i|=0\;,\quad |u|>0\;,   
\end{equation}
for the product $(x\bullet u)$ to be covariant. Notice that the structures on the right hand side above are completely fixed by degree and symmetry, and the assumption is actually on the relative coefficients. 

\subsection{Relation with differential graded Lie algebras}

Looking at the properties \eqref{motivation4} and \eqref{motivation5} one immediately notices the resemblance with the graded Jacobi identity and graded Leibniz rule of differential graded Lie algebras. In fact, one can show that this is precisely the case upon suspension, \emph{i.e.} degree shifting, of the graded vector spaces $X_n\,$. To start with, we shall define the degree shifted vector space $\widetilde X=\bigoplus_{n=1}^\infty\widetilde X_n$  and the suspension $s$ by
\begin{equation}
\begin{split}
&s:X_n\rightarrow \widetilde X_{n+1}  \;,\quad \tilde a:=s a  \;,\\
& |\tilde a|=|a|+1\;.
\end{split}    
\end{equation}
The bullet product on $X$ translates, upon suspension, to a graded antisymmetric bracket on $\widetilde X$ defined by
\begin{equation}\label{dgla bracket}
[\tilde a,\tilde b]:=(-1)^{|a|+1}s(a\bullet b)    \;,
\end{equation}
which should not be confused with the antisymmetrization of the Leibniz product. The differential $\cD$ can also be defined on the suspended spaces by $\cD\tilde a:=s\cD a\,$. Degree counting shows that the differential, that we still denote by $\cD\,$, retains degree $-1\,$, while the bracket \eqref{dgla bracket} has degree zero.

With these definitions one can show that properties \eqref{motivation4} and \eqref{motivation5} become the usual graded Jacobi identity for the bracket \eqref{dgla bracket} and graded Leibniz compatibility of the differential, namely
\begin{equation}\label{dgla relations}
\begin{split}
&[[\tilde a,\tilde b],\tilde c]+(-1)^{|\tilde a|(|\tilde b|+|\tilde c|)}[[\tilde b,\tilde c],\tilde a]+(-1)^{|\tilde c|(|\tilde a|+|\tilde b|)}[[\tilde c,\tilde a],\tilde b]=0   \;,\\
& \cD[\tilde a,\tilde b]=[\cD\tilde a,\tilde b]+(-1)^{|\tilde a|}[\tilde a,\cD\tilde b]\;,\quad |\tilde a|\,,|\tilde b|>1\;.
\end{split}    
\end{equation}
Finally, the original Leibniz algebra $(X_0,\circ)$ can be transported to $\widetilde X_1$ by 
\begin{equation}
\tilde x\circ\tilde y :=s(x\circ y)  \;, 
\end{equation}
so that the Leibniz property is unchanged. The suspended Leibniz product has intrinsic degree $-1\,$, and indeed closes on $\widetilde X_1\,$.

Even though the differential graded Lie algebra (dgLa) structure appearing in \eqref{dgla relations} allows for a more familiar interpretation of the properties \eqref{motivation4} and \eqref{motivation5}, one should keep in mind that the Leibniz product $\circ$ is an independent algebraic structure. In particular, its compatibility properties with the differential and the dgLa bracket, given by the suspended version of \eqref{motivation-1}, \eqref{motivation0}, \eqref{motivation1} and \eqref{motivation6}, need to be imposed as separate requirements.

This construction simplifies considerably if one extends the graded vector space $\widetilde X$ to non-positive degrees and declares the differential and the bracket to be defined on the whole space, thus making $(\widetilde X,\,[\,, ],\,\cD)$ into a differential graded Lie algebra. In particular, the newly introduced space $\widetilde X_0=s\, X_{-1}$ is a genuine Lie algebra $\mathfrak{g}$ and the dgLa bracket gives to all graded spaces $\widetilde X_n$ a $\mathfrak{g}-$module structure via
\begin{equation}
\rho_{\tilde t}\,\tilde a:=[\tilde t,\tilde a]\;,\quad \tilde t\in \widetilde X_0\,,\;\tilde a\in\widetilde X\;.   
\end{equation}
The crucial difference of the extended space formulation is that the differential can now act on the Leibniz space $\widetilde X_1=s\,X_0\,$, yielding a map
\begin{equation}
\cD: \widetilde X_1\rightarrow\mathfrak{g}    
\end{equation}
that can be interpreted as an abstract embedding tensor, in the language of gauged supergravity. This in turn allows one to \emph{define} the Leibniz product as a derived one via
\begin{equation}\label{Leibniz defined}
\tilde x\circ\tilde y:=-[\cD\tilde x,\tilde y] \;,   
\end{equation}
or, before suspension, as $x\circ y:=-\cD x\bullet y\,$. The graded Jacobi identity and compatibility of $\cD$ are sufficient to prove that the product \eqref{Leibniz defined} does indeed obey the Leibniz property. Moreover, the Lie derivative is also universally defined by
\begin{equation}
\cL_{\tilde x}\tilde a:=-[\cD \tilde x,\tilde a] \;,\quad \tilde x\in \widetilde X_1\,,\;\tilde a\in\widetilde X\;.    
\end{equation}
Finally, all the compatibility conditions \eqref{motivation-1}, \eqref{motivation0}, \eqref{motivation1} and \eqref{motivation6} are ensured by the dgLa structure of the entire vector space, requiring only the graded Jacobi identity and compatibility of the differential.

At this point it is evident  that the dgLa structure on the unbounded space allows for a more concise construction. However, it should be noted that assuming the existence of the Lie algebra $\mathfrak{g}$ and its action on $\widetilde X$ is not needed in order to construct the tensor hierarchy and is a considerable piece of extra data to be given as input. Moreover, so far there is no clear field theoretic interpretation of the spaces in negative degrees. Nevertheless, given any Leibniz algebra $(X_0,\circ)$ it is always possible to define an associated Lie algebra by modding out the symmetric part of the Leibniz product. This yields a Lie algebra $\mathfrak{g}_0\sim X_0/\cD(X_1)$ that, however, almost never coincides with the Lie algebra $\mathfrak{g}=s \,X_{-1}\,$, but is rather a subalgebra of the latter. For instance, in gauged supergravity the underlying Lie algebra $\mathfrak{g}$ is given by the global symmetry algebra of the ungauged phase, while in double and exceptional field theory such an algebra  was identified only recently \cite{Hohm:2018ybo,Hohm:2019wql}. This shows that augmenting the original graded vector space $X$ to negative degrees and extending the action of the differential and the bullet requires a significant amount of extra structure, that is not  necessary in order to define a consistent tensor hierarchy.
For this reason, in the present paper we shall investigate tensor hierarchies in terms of the most minimal set of algebraic structures that guarantee their consistency. We will consider the graded vector space $X$ concentrated in non-negative degrees, with the Leibniz algebra $(X_0,\circ)$ as a fundamental structure. As discussed in the present section, the consistency relations \eqref{motivation-1}, \eqref{motivation0}, \eqref{motivation1} and \eqref{motivation6} cannot be derived, in this case, by more fundamental ones, and we shall demand them as additional axioms that will be collected in the next subsection.

\subsection{Axioms}

We are now ready to provide the list of structures and axioms defining what we name an infinity-enhanced Leibniz algebra. In this section we will prove that the given axioms allow one to define a generalized Lie derivative that acts covariantly on all  algebraic structures, and closes on itself modulo trivial transformations. 
\\
An infinity-enhanced Leibniz algebra consists of the quadruple $(X,\circ,\cD,\bullet)\,$. $X$ is an $\mathbb{N}$-graded vector space, where sometimes 
we single out the degree zero subspace:
\begin{equation}
X=\bigoplus_{n=0}^\infty X_n=X_0\oplus \bar X\;.    
\end{equation}
$X_0$ is endowed with a (left) Leibniz product $\circ:X_0\otimes X_0\to X_0\,$, obeying
\begin{equation}
x\circ(y\circ z)=(x\circ y)\circ z+y\circ(x\circ z)\;. \end{equation}
$\cD$ is a degree $-1$ differential acting on $\bar X\,$:
\begin{equation}
...\longrightarrow\; X_n\;\stackrel{\cD}{\longrightarrow}\;X_{n-1} \;... \stackrel{\cD}{\longrightarrow}\;X_1\;\stackrel{\cD}{\longrightarrow}\;X_0\;,  \qquad\cD^2 \ = \ 0\;,
\end{equation}
and $\bullet$ is a graded commutative product of degree $+1$ defined on the whole space $X\,$:
\begin{equation}
\bullet:X_i\otimes X_j\to X_{i+j+1}\;,\quad a\bullet b=(-1)^{|a||b|}b\bullet a\;.    
\end{equation}
This quadruple defines an infinity-enhanced Leibniz algebra provided 
\begin{equation}\label{axioms}
\begin{split}
1) \quad &\cD u\circ x=0\;,\quad\forall \;u\in X_1,\; x\in X_0 \;, \\
2)\quad & \cD(x\bullet y)=x\circ y+y\circ x\;,\quad\forall \;x,y\in X_0\;,\\
3)\quad &\cD(x\bullet(y\bullet z))=(x\circ y)\bullet z+(x\circ z)\bullet y-(y\circ z+z\circ y)\bullet x\;,\quad\forall\;x,y,z\in X_0\;,\\
4)\quad &\cD(x_{[1}\bullet(x_{2]}\bullet u))=2\,x_{[2}\bullet\cD(x_{1]}\bullet u)+x_{[2}\bullet(x_{1]}\bullet\cD u)+[x_1,x_2]\bullet u\,,\;\;\forall\,x_1,x_2\in X_0,u\in \bar X\;,\\
5) \quad&\cD(u\bullet v)+\cD u\bullet v+(-1)^{|u|}u\bullet\cD v=0\;,\quad\forall\;u,v\in\bar X\;,\\
6)\quad& (-1)^{|a|}a\bullet(b\bullet c)+(a\bullet b)\bullet c+(-1)^{|b||c|}(a\bullet c)\bullet b=0\;,\quad\forall\;a,b,c\in X\;. 
\end{split}    
\end{equation}
The generalized Lie derivative is defined as
\begin{equation}
\begin{split}
&\cL_x y:=x\circ y\;,\quad\forall\; x,y\in X_0\;,\\
&\cL_x u:=x\bullet\cD u+\cD(x\bullet u)\;,\quad\forall \;x\in X_0\;, u\in \bar X\;.
\end{split}    
\end{equation}
From axioms 1) and 5) it is immediate that $\cD$-exact degree zero elements generate trivial Lie derivatives, \emph{i.e.}
\begin{equation}
\cL_{\cD u}a=0\;,\quad\forall\;u\in X_1\;,\,a\in X\;.    
\end{equation}

\paragraph{Covariance}

As the first statement of covariance, we see that the Lie derivative commutes with the differential $\cD\,$,
\begin{equation}\label{cDLiecommute}
\big[\cL_x,\cD\big]=0\;,\quad \forall\;x\in X_0 \;.  
\end{equation}
This is obvious by construction when the commutator acts on an element $u$ with $|u|>1\,$, since then 
$\cL_x=\iota_x\cD+\cD\iota_x$, c.f.~(\ref{DcalLCommute}). 
For $|u|=1$ one has
\begin{equation}
\big[\cL_x,\cD\big]u=x\circ\cD u-\cD(x\bullet\cD u+\cD(x\bullet u))=x\circ\cD u-\cD(x\bullet\cD u)=0\;, \end{equation}
upon using 1) and 2).
Covariance of the Leibniz product $\circ$ itself is just a rewriting of its defining property:
\begin{equation}
\cL_x(y\circ z)=(\cL_x y)\circ z+y\circ(\cL_xz) \;.   
\end{equation}
Covariance of the bullet product, \emph{i.e.}
\begin{equation}\label{bulletcovariance}
\cL_x(a\bullet b)=(\cL_xa)\bullet b+a\bullet(\cL_xb)\;,\quad \forall\;x\in X_0\;,\;a,b\in X\;,    
\end{equation}
has to be proved in different steps, depending on the degrees of $a$ and $b\,$. For $a$ and $b$ in $X_0$ one has
\begin{equation}
\begin{split}
\cL_x(a\bullet b)&=x\bullet\cD(a\bullet b)+\cD(x\bullet(a\bullet b))=x\bullet(a\circ b+b\circ a)+\cD(x\bullet(a\bullet b))  \\
&=(x\circ a)\bullet b+(x\circ b)\bullet a=(\cL_xa)\bullet b+a\bullet(\cL_xb)
\end{split}    
\end{equation}
thanks to 3).
In the case $a\in X_0$ and $b\in\bar X$ we have to use properties 4), 5) and 6) (in the proof we rename $a=y$ to make clear that it has zero degree):
\begin{equation}
\begin{split}
&\cL_x(y\bullet b)=x\bullet\cD(y\bullet b)+\cD(x\bullet(y\bullet b))\\
&=x\bullet\cD(y\bullet b)+\tfrac12\,\Big[\cD(x\bullet(y\bullet b))+\cD(y\bullet(x\bullet b))\Big]+\tfrac12\,\Big[\cD(x\bullet(y\bullet b))-\cD(y\bullet(x\bullet b))\Big]\\
&=-\tfrac12\,\cD((x\bullet y)\bullet b)+y\bullet\cD(x\bullet b)+\tfrac12\,y\bullet(x\bullet\cD b)-\tfrac12\,x\bullet(y\bullet\cD b)+[x,y]\bullet b\\
&=\{x,y\}\bullet b-\tfrac12\,(x\bullet y)\bullet\cD b+y\bullet\cL_xb-\tfrac12\,y\bullet(x\bullet\cD b)-\tfrac12\,x\bullet(y\bullet\cD b)+[x,y]\bullet b\\
&=(x\circ y)\bullet b+y\bullet\cL_xb\;.
\end{split}    
\end{equation}
For both $a$ and $b$ in $\bar X$ one uses repeatedly properties 5) and 6) to get
\begin{equation}
\begin{split}
&\cL_x(a\bullet b)=x\bullet\cD(a\bullet b)+\cD(x\bullet(a\bullet b))\\
&= -x\bullet[\cD a\bullet b+(-1)^{|a|}a\bullet\cD b]-\cD[(x\bullet a)\bullet b+(-1)^{|a|}a\bullet(x\bullet b)]\\
&=(x\bullet\cD a)\bullet b-(-1)^{|a|}\cD a\bullet(x\bullet b)+(-1)^{|a|}(x\bullet a)\bullet\cD b+a\bullet(x\bullet\cD b)\\
&\;\quad-\cD[(x\bullet a)\bullet b+(-1)^{|a|}a\bullet(x\bullet b)]\\
&=(\cL_x a)\bullet b-\cD(x\bullet a)\bullet b-(-1)^{|a|}\cD a\bullet(x\bullet b)+(-1)^{|a|}(x\bullet a)\bullet\cD b+a\bullet(\cL_x b)-a\bullet\cD(x\bullet b)\\
&\;\quad-\cD[(x\bullet a)\bullet b+(-1)^{|a|}a\bullet(x\bullet b)]\\
&=(\cL_x a)\bullet b+a\bullet(\cL_x b)\;,
\end{split}    
\end{equation}
thus proving covariance \eqref{bulletcovariance} for arbitrary elements $a,b\in X\,$.

\paragraph{Closure}

Since the Lie derivative will be used to define symmetry variations and covariant derivatives, we have to show that it closes under commutation, namely
\begin{equation}\label{Lieclosure}
\big[\cL_x,\cL_y\big]\,a=\cL_{[x,y]}\,a\;.    
\end{equation}
When acting on a degree zero element, this is ensured by the Leibniz property of $\circ\,$, 
as displayed in (\ref{gaugealgebra0}), 
which yields  \eqref{Lieclosure} together with the triviality property $\cL_{\{x,y\}}=0\,$.
In order to prove closure on higher degree elements we need the covariance properties  \eqref{cDLiecommute} and \eqref{bulletcovariance}  that have just been proven:
\begin{equation}
\begin{split}
\cL_x\cL_ya&=\cL_x\Big[y\bullet\cD a+\cD(y\bullet a)\Big] \\
&=(x\circ y)\bullet\cD a+y\bullet\cD(\cL_xa)+\cD\Big[(x\circ y)\bullet a+y\bullet(\cL_xa)\Big]\\
&=\cL_{x\circ y}a+\cL_y\cL_xa\;.
\end{split}    
\end{equation}

We close this section by giving a brief summary of where 
the algebraic structures axiomatized here appear in physical models. For instance, in gauged supergravity (see \emph{e.g.} \cite{deWit:2008ta}) the vector fields take values in some representation $R_1$ of the Lie algebra $\mathfrak{g}$ of global symmetries of the ungauged phase. The Leibniz space $X_0$ is then identified with the representation space $R_1$,  and the so called embedding tensor map $\vartheta: R_1\to \mathfrak{g}$ allows one  to define the Leibniz product as $x\circ y=\rho_{\vartheta(x)}y\,$ for $x,y\in R_1\,$. The first bullet product $\bullet$ is then defined as a projection from the tensor product of two $R_1$ representations to the $R_2\equiv X_1$ representation carried by the two-forms: $\bullet:R_1\otimes R_1\to R_2\,$, and so on. 
In double field theory \cite{Hohm:2019wql} and exceptional field theory \cite{Hohm:2015xna,Hohm:2019wql,Hohm:2018ybo}  the Leibniz product is given in terms of generalized Lie derivatives, and both the bullet product $\bullet$ and differential $\cD$ are given by algebraic and/or differential operators acting on the internal space. Similarly, for a 3D gauge theory based on an infinite-dimensional Leibniz algebra see \cite{Hohm:2018git}.

\section{Exact tensor hierarchy}\label{Sec:TensorH}

The main goal of this section is to show that the infinity-enhanced Leibniz algebra, defined in the previous section by the set of axioms \eqref{axioms}, allows us to construct the tensor hierarchy to all orders. In particular, we will show that it is possible to define gauge covariant curvatures $\cF_{p+1}$
for $p$-form gauge fields $A_p$ of arbitrary degree. 
Consistency of the tensor hierarchy is established by showing that the curvatures obey a set of Bianchi identities. Indirectly, 
this establishes gauge covariance.

\subsection{General strategy} 
We start by briefly outlining the general strategy:  
When constructing the gauge theory step by step, as in section \ref{sec:THlowlevels}, one starts from the one-form\footnote{From now on, having to deal with forms of arbitrary degree, we use intrinsic differential form notation with normalization $A_p=\tfrac{1}{p!}\,A_{\mu_1...\mu_p}\,dx^{\mu_1}...dx^{\mu_p}\,$, and different forms are only named by their degree, so that $A_{2\,\mu\nu}=B_{\mu\nu}$ of section \ref{sec:THlowlevels} and so on. Moreover, we use a slightly different normalization for the $\Omega_3$ Chern-Simons form compared to section \ref{sec:THlowlevels}, in order to avoid cluttering formulas with factorial coefficients.} $A_1=A_\mu\,dx^\mu\,$, taking values in the Leibniz algebra $X_0\,$,
and postulates the gauge transformation\footnote{As shown in section \ref{sec:THlowlevels} the $\cD$-exact part of the gauge transformation is required by closure.} $\delta A_1=D\lambda_0-\cD\lambda_1\,$, where the covariant derivative is defined as
\begin{equation}
D=dx^\mu\,D_\mu:=dx^\mu\,(\del_\mu-\cL_{A_\mu}) \;.   
\end{equation}
The naive Yang-Mills curvature $F_2=dA_1-\tfrac12 A_1\circ A_1$
transforms covariantly only modulo a $\cD$-exact term: $\delta F_2=\cL_{\lambda_0}F_2+\cD(...)\,$, forcing us to introduce an $X_1$-valued two-form $A_2$ whose gauge transformation can be adjusted to make the full two-form curvature $\cF_2=F_2+\cD A_2$ covariant, \emph{i.e.} $\delta\cF_2=\cL_{\lambda_0}\cF_2\,$. At this point, the most efficient way to determine the three-form curvature is to take the covariant curl of $\cF_2\,$, yielding the Bianchi identity
$
D\cF_2=\cD F_3    
$
with 
\begin{equation}\label{firstCS}
\begin{split}
F_3&=DA_2+\Omega_3\;,\\
\Omega_3&=-\tfrac12\,A_1\bullet dA_1+\tfrac16\,A_1\bullet(A_1\circ A_1)\;.
\end{split}    
\end{equation}
As for the lower order, the curvature $F_3$ is covariant only modulo $\cD$-exact terms: $\delta F_3=\cL_{\lambda_0}F_3+\cD(...)\,$. Again, this can be cured by introducing an $X_2$-valued gauge potential $A_3$ and fixing its gauge transformation so that $\cF_3=F_3+\cD A_3$ transforms covariantly. The procedure continues in the same way up to the top form, for a given spacetime dimension, but it becomes very cumbersome quite quickly, as can be appreciated from the first steps carried out explicitly in section \ref{sec:THlowlevels}.\\
From the examples at low form degree it is clear that at every order the Bianchi identities alone fix the next order curvature up to a $\cD$-exact term. We aim thus at finding curvatures $\cF_p=DA_{p-1}+\cD A_p+...$ that obey Bianchi identities among themselves for arbitrary form degrees. 

In order to do so, we shall first focus on the Chern-Simons-like terms such as the $\Omega_3$ in \eqref{firstCS}. Such composite $p$-forms, entirely built out of the one-form $A_1\,$, appear at every order in a brute-force calculation of the curvatures: $\cF_p=DA_{p-1}+\cD A_p+\Omega_p(A_1)+\cdots$.  Perhaps counter-intuitively, it is the term $\Omega_p(A_1)\,$, rather than  $DA_{p-1}\,$, that helps us to determine the all-order structure of $\cF_p\,$ since, as we shall prove in appendix \ref{App:CSproof} , these pseudo-Chern-Simons forms already obey Bianchi identities. We thus define the pseudo Chern-Simons (CS) $n$-form by
\begin{equation}
\Omega_n(A)=\tfrac{(-1)^n}{(n-1)!}\,(\iota_A)^{n-2}\big[dA-\tfrac{1}{n}\,A\circ A\big]       \;,\quad |\Omega_n|=n-2\;,
\end{equation}
where $A\equiv A_1$ and we have introduced the operator
\begin{equation}
\iota_A x:= A\bullet x \;.  
\end{equation}
The form index $n$ is related to the powers of $A$ as $\Omega_n\sim A^{n-2}dA+A^n$ such that, for instance, one has
$\Omega_3=-\tfrac{1}{2}\,A\bullet dA+\tfrac16\,A\bullet(A\circ A)\;,$ $\Omega_4=\tfrac16\,A\bullet(A\bullet dA)-\tfrac{1}{24}\,A\bullet(A\bullet(A\circ A))$ and so on. It will be convenient to include the pure Yang-Mills curvature $F_2$ in this family as $\Omega_2=dA-\tfrac12\,A\circ A\equiv F_2\,$.
Notice that the $\Omega_n$  have form degree $n$ and intrinsic degree $n-2\,$, making them $\bullet$-commutative, \emph{i.e.}
\begin{equation}
\Omega_k\bullet\Omega_l=\Omega_l\bullet\Omega_k\;,\quad \forall\;k,l\geq2  \;.
\end{equation}
Moreover, from assumption 6) of \eqref{axioms}, upon combining internal and form degrees, one can derive
\begin{equation}
-\iota_A(\Omega_k\bullet\Omega_l)=(\iota_A\Omega_k)\bullet\Omega_l+\Omega_k\bullet(\iota_A\Omega_l)\;.     
\end{equation}
The crucial property of this family of differential forms is that it closes (quadratically) under the action of the covariant derivative $D=d-\cL_A\,$:
\begin{equation}\label{CSrelations}
D\Omega_{n}+\tfrac12\sum_{k=2}^{n-1}\Omega_{k}\bullet\Omega_{n+1-k}=\cD\Omega_{n+1}\;,\quad n\geq 2\;.    
\end{equation}
We will prove \eqref{CSrelations} by induction in appendix \ref{App:CSproof}. 

\subsection{Curvatures and Bianchi identities to all orders}

In this section we are going to work with arbitrary $p$-form gauge potentials $A_p\,$, defined as differential $p$-forms over a spacetime manifold $M\,$, taking values in $X_{p-1}\,$, namely
\begin{equation}
A_p:=\tfrac{1}{p!}\,A_{\mu_1...\mu_p}(x)\,dx^{\mu_1}\cdots dx^{\mu_p}\;\in\,\Omega^p(M)\otimes X_{p-1}\;,    
\end{equation}
where from now on we shall omit the symbol for wedge products.

We are now ready to introduce curvatures $\cF_{p+1}\in \Omega^{p+1}(M)\otimes X_{p-1}$ for arbitrary $p$-form gauge fields as
\begin{equation}\label{Fallorders}
\cF_{p+1}=\sum_{k_i\geq2}^{p}\sum_{N=1}^{\scaleobj{0.6}{\left[\tfrac{p}{2}\right]}}\tfrac{(-1)^{N-1}}{N!} \iota_{k_1}...\iota_{k_{N-1}}DA_{k_N}+
\sum_{k_i\geq2}^{p+1}\sum_{N=1}^{\scaleobj{0.6}{\left[\tfrac{p+1}{2}\right]}}\tfrac{(-1)^{N-1}}{N!} \iota_{k_1}...\iota_{k_{N-1}}\Big(\cD A_{k_N}+N\,\Omega_{k_N}\Big)\;,
\end{equation}
where $A_p$ with $p\geq2$ are the higher form gauge fields,
\begin{equation}
\Omega_p=\tfrac{(-1)^p}{(p-1)!}\,\iota_1^{p-2}\Big(dA_1-\tfrac1n\,A_1\circ A_1\Big)  \;,\quad p\geq2  
\end{equation}
are the pseudo CS forms just introduced,
and we use the shorthand notation
\begin{equation}
\iota_k\,x:= A_k\bullet x    \;.
\end{equation}
The sums $\sum_{k_i\geq2}^n$ run over all possible $\{k_1,...,k_N\}$ with ${k_i\geq2}$, constrained by ${\sum_{i=1}^Nk_i=n}\,$. We will prove that the curvatures defined in \eqref{Fallorders} obey the Bianchi identities
\begin{equation}\label{Bianchiallorders}
D\cF_n+\tfrac12\sum_{k=2}^{n-1}\cF_k\bullet\cF_{n+1-k}=\cD\cF_{n+1} \;,\quad n\geq2 \;,  
\end{equation}
that are the benchmark for proving recursively the gauge covariance of the $\cF_n$'s.
Before giving the proof of \eqref{Bianchiallorders} for arbitrary form degree, we spend a few words to justify the form of \eqref{Fallorders}. When constructing the tensor hierarchy step by step, starting from the one-form, one is led to identify the first few gauge covariant curvatures as
\begin{equation}\label{firstFs}
\begin{split}
\cF_2&= dA_1-\tfrac12\,A_1\circ A_1+\cD A_2=F_2+\cD A_2\equiv \Omega_2+\cD A_2\;,\\
\cF_3&= DA_2+\Omega_3+\cD A_3\;,\\
\cF_4&=DA_3+\Omega_4-A_2\bullet\Omega_2-\tfrac12\,A_2\bullet\cD A_2+\cD A_4\;,
\end{split}    
\end{equation}
that immediately suggest  to consider
\begin{equation}\label{Flinearansatz}
\cF_n=DA_{n-1}+\Omega_n+\cD A_n+\cdots 
\end{equation}
as a starting point for the curvature, which indeed coincides with the ${N=1}$ term of \eqref{Fallorders}. When considering the $DA_{n-1}$ term, we recall that $DA_1$ is not defined (the pure Yang-Mills part of $\cF_2$ is $\Omega_2$) and we formally set $DA_1\equiv0$ when necessary. In order to guess the structure of the extra terms in \eqref{Flinearansatz} it is useful to push the step by step procedure a bit further to find
\begin{equation}
\cF_5=DA_4+\Omega_5-A_2\bullet\Omega_3-A_3\bullet\Omega_2-\tfrac12\,A_2\bullet DA_2-\alpha\,A_2\bullet\cD A_3+(\alpha-1)\,A_3\bullet\cD A_2+\cD A_5  \;.  
\end{equation}
The ambiguity, parametrized by $\alpha\in\mathbb{R}\,$, amounts to a field redefinition ${A_5\to A_5+A_2\bullet A_3}\,$, that we fix by choosing the symmetric point $\alpha=\tfrac12\,$. One can easily see that, starting from $\cF_5\,$, analogous ambiguities show up at every level, due to possible field redefinitions of the $\cD$-exact term $\cD A_n$ in \eqref{Flinearansatz}. We choose the symmetric point for all of them by demanding that the coefficients, as displayed in \eqref{Fallorders}, do not depend on the set $\{k_i\}_{i=1}^N$ of form degrees.
\\
In order to formulate the ansatz for $\cF_n$ we define a new degree, that we name twist, given by the difference between form degree and internal degree of a given field. We recall that gauge fields $A_p$ have internal degree $p-1\,$, the differential $\cD$ has internal degree $-1\,$, and the products $\circ$ and $\bullet$ have degree zero and $+1\,$, respectively. It is thus clear that all gauge fields have twist $+1\,$, while all curvatures have twist $+2$ and the operator $\iota_k$ has twist zero. Therefore, the only way to make a twist $+2$ object from a higher gauge field $A_p$ with ${p\geq2}$ is to act with an arbitrary number of $\iota_{k_i}$ on the building blocks $\cD A_p$ and $DA_p\,$.\footnote{In principle one can consider $dA_p$ and $\cL_{A_1}A_p$ separately, but there is no point in breaking the covariant derivative.}  As for the vector $A_1\,$, a basis of twist $+2$ forms can be constructed by acting with an arbitrary number of $\iota_{k_i}$ on the building block\footnote{As before, one could consider $dA_1$ and $A_1\circ A_1$ separately, but there is no advantage in breaking up $F_2=\Omega_2\,$.} $\Omega_p\,$, finally leading to the ansatz
\begin{equation}\label{Fansatz}
\cF_n=DA_{n-1}+\Omega_n+\cD A_n+\sum_{k_i\geq2}^n\sum_{N=2}^{\scaleobj{0.6}{\left[\tfrac{n}{2}\right]}}\iota_{k_1}...\iota_{k_{N-1}}\Big(\alpha_N\,DA_{k_N-1}+\beta_N\,\cD A_{k_N}+\gamma_N\,\Omega_{k_N}\Big) \;,   
\end{equation}
where the leading terms can be included as the ${N=1}$ part of the sum for the initial values $\alpha_1=\beta_1=\gamma_1=1\,$.
Despite the usual point of view of seeing $DA_{n-1}$ as the leading term of the field strength $\cF_n\,$, it is more convenient to split it as
\begin{equation}
\cF_n=\Omega_n+\Delta\cF_n    
\end{equation}
in order to prove the Bianchi identities, since the pseudo CS forms already obey \eqref{CSrelations}, yielding
\begin{equation}
\begin{split}
D\cF_n&=D\Omega_n+D\Delta\cF_n=-\tfrac12\sum_{k=2}^{n-1}\Omega_k\bullet\Omega_{n+1-k}+\cD\Omega_{n+1}+D\Delta\cF_n   \\
&= -\tfrac12\sum_{k=2}^{n-1}\cF_k\bullet\cF_{n+1-k}+\cD\cF_{n+1}\\
&+\sum_{k=2}^{n-1}\big[\Omega_k\bullet\Delta\cF_{n+1-k}+\tfrac12\,\Delta\cF_k\bullet\Delta\cF_{n+1-k}\big]+D\Delta\cF_n-\cD\Delta\cF_{n+1}\;.
\end{split}    
\end{equation}
Proving the Bianchi identity \eqref{Bianchiallorders} thus amounts to showing that
\begin{equation}\label{Bianchiproof1}
\cD\Delta\cF_{n+1}-D\Delta\cF_n=    \sum_{k=2}^{n-1}\big[\Omega_k\bullet\Delta\cF_{n+1-k}+\tfrac12\,\Delta\cF_k\bullet\Delta\cF_{n+1-k}\big]\;,
\end{equation}
where 
\begin{equation}
\begin{split}
\Delta\cF_n&=f_n^A+f_n^\Omega\;,\quad\text{with}\\
f_n^A&:=\sum_{k_i\geq2}^{n-1}\sum_{N\geq1}\alpha_N\,\iota_{k_1}...\iota_{k_{N-1}}D A_{k_N}+\sum_{k_i\geq2}^n\sum_{N\geq1}\beta_N\,\iota_{k_1}...\iota_{k_{N-1}}\cD A_{k_N}\;,\\
f_n^\Omega&:=\sum_{k_i\geq2}^n\sum_{N\geq2}\gamma_N\,\iota_{k_1}...\iota_{k_{N-1}}\Omega_{k_N}\;.
\end{split}    
\end{equation}
Let us first focus on the term
$Df_n^A-\cD f_{n+1}^A\,$,
for which we have to treat objects of the form (at fixed $N$)
\begin{equation}
\sum_{k_i\geq2} D[\iota_{k_1}...\iota_{k_{N-1}}B_{k_N}]\;,\quad   \sum_{k_i\geq2} \cD[\iota_{k_1}...\iota_{k_{N-1}}B_{k_N}]\;,
\end{equation}
where $B_k$ is a $k$-form of twist $+2\,$.
By using the graded Leibniz rule of the covariant derivative one has
\begin{equation}
\begin{split}
D[\iota_{k_1}...\iota_{k_{N-1}}B_{k_N}]&=\sum_{l=0}^{N-2}(-1)^{k_1+...+k_l}\iota_{k_1}...\iota_{k_l}[DA_{k_{l+1}}\bullet(\iota_{k_{l+2}}...\iota_{k_{N-1}}B_{k_N})]\\
&+(-1)^{k_1+...+k_{N-1}}\iota_{k_1}...\iota_{k_{N-1}}DB_{k_N}  \;. 
\end{split}
\end{equation}
The term in square brackets can be further manipulated by recursively using the identity
\begin{equation}
(-1)^k\iota_k(f\bullet g)=(\iota_kf)\bullet g+f\bullet(\iota_k g)\;,
\end{equation}
that is axiom 6) of \eqref{axioms} for $f$ and $g$ with even twist, to get
\begin{equation}\label{Pascaltriangle}
\begin{split}
\sum_{k_i\geq2}\sum_{l=0}^{N-2}&(-1)^{k_1+...+k_l}\iota_{k_1}...\iota_{k_l}[DA_{k_{l+1}}\bullet(\iota_{k_{l+2}}...\iota_{k_{N-1}}B_{k_N})]\\
&=\sum_{k_i\geq2}\sum_{l=0}^{N-2}\sum_{m=0}^l\binom{l}{m}\,(\iota_{k_1}...\iota_{k_m}DA_{k_{m+1}})\bullet(\iota_{k_{m+2}}...\iota_{k_{N-1}}B_{k_N})\\
&=\sum_{k_i\geq2}\sum_{m=0}^{N-2}\Big[\sum_{l=m}^{N-2}\binom{l}{m}\Big]\,(\iota_{k_1}...\iota_{k_m}DA_{k_{m+1}})\bullet(\iota_{k_{m+2}}...\iota_{k_{N-1}}B_{k_N}) \\
&=\sum_{k_i\geq2}\sum_{m=0}^{N-2}\binom{N-1}{m+1}\,(\iota_{k_1}...\iota_{k_m}DA_{k_{m+1}})\bullet(\iota_{k_{m+2}}...\iota_{k_{N-1}}B_{k_N}) \;,
\end{split}    
\end{equation}
finally giving
\begin{equation}\label{Didentity}
\begin{split}
\sum_{k_i\geq2}D[\iota_{k_1}...\iota_{k_{N-1}}B_{k_N}]=\sum_{k_i\geq2}&\Big\{\sum_{m=0}^{N-2}\binom{N-1}{m+1}\,(\iota_{k_1}...\iota_{k_m}DA_{k_{m+1}})\bullet(\iota_{k_{m+2}}...\iota_{k_{N-1}}B_{k_N})\\
&+(-1)^{k_1+...+k_{N-1}}\iota_{k_1}...\iota_{k_{N-1}}DB_{k_N}\Big\}\;.
\end{split}
\end{equation}
In an almost identical way, by using 5) of \eqref{axioms}, one proves that
\begin{equation}\label{cDidentity}
\begin{split}
-\sum_{k_i\geq2}\cD[\iota_{k_1}...\iota_{k_{N-1}}B_{k_N}]=\sum_{k_i\geq2}&\Big\{\sum_{m=0}^{N-2}\binom{N-1}{m+1}\,(\iota_{k_1}...\iota_{k_m}\cD A_{k_{m+1}})\bullet(\iota_{k_{m+2}}...\iota_{k_{N-1}}B_{k_N})\\
&-(-1)^{k_1+...+k_{N-1}}\iota_{k_1}...\iota_{k_{N-1}}\cD B_{k_N}\Big\}\;.
\end{split}
\end{equation}
Using the identities \eqref{Didentity} and \eqref{cDidentity} on $Df_n^A-\cD f_{n+1}^A\,$ one obtains
\begin{equation}\label{bigD-cDidentity1}
\begin{split}
&Df_n^A-\cD f_{n+1}^A=\sum_{k_i\geq2}^n\sum_{N\geq1}(\beta_N-\alpha_N)(-1)^{k_1+...+k_{N-1}}\iota_{k_1}...\iota_{k_{N-1}}D\cD A_{k_N}\\
&+\sum_{k_i\geq2}^{n}\sum_{N\geq2}\sum_{m=0}^{N-2}\Big[\beta_N\,\binom{N-1}{m+1}+\alpha_N\,\binom{N-1}{N-1-m}\Big](\iota_{k_1}...\iota_{k_m}DA_{k_{m+1}})\bullet(\iota_{k_{m+2}}...\iota_{k_{N-1}}\cD A_{k_N})\\
&+\sum_{k_i\geq2}^{n-1}\sum_{N\geq2}\alpha_N\sum_{m=0}^{N-2}\binom{N-1}{m+1}(\iota_{k_1}...\iota_{k_m}DA_{k_{m+1}})\bullet(\iota_{k_{m+2}}...\iota_{k_{N-1}}DA_{k_N})\\
&+\sum_{k_i\geq2}^{n+1}\sum_{N\geq2}\beta_N\sum_{m=0}^{N-2}\binom{N-1}{m+1}(\iota_{k_1}...\iota_{k_m}\cD A_{k_{m+1}})\bullet(\iota_{k_{m+2}}...\iota_{k_{N-1}}\cD A_{k_N})\\
&-\sum_{k_i\geq2}^{n-1}\sum_{N\geq1}\alpha_N\,(-1)^{k_1+...+k_{N-1}}\iota_{k_1}...\iota_{k_{N-1}}\cL_{\Omega_2}A_{k_N}\;,
\end{split}    
\end{equation}
where we have used $D^2=-\cL_{\Omega_2}\,$.
By looking at the right hand side of \eqref{Bianchiproof1}
one sees that there are no terms containing $D\cD A_k\,$, hence fixing $\beta_N=\alpha_N$ from the first term of \eqref{bigD-cDidentity1}.
Furthermore, by looking at the diagonal terms of the form $(\iota^mDA)\bullet(\iota^{N-2-m}DA)$ and $(\iota^m\cD A)\bullet(\iota^{N-2-m}\cD A)\,$, one notices that they are manifestly symmetric (under the $\sum_{k_i}$) upon the exchange $m\to N-2-m\,$, thereby projecting the corresponding coefficients to their manifest symmetric part:
\begin{equation*}
\binom{N-1}{m+1}\rightarrow\frac12\,\Big[\binom{N-1}{m+1}+\binom{N-1}{N-1-m}\Big]=\frac12\,\Big[\binom{N-1}{m+1}+\binom{N-1}{m}\Big]=\frac12\,\binom{N}{m+1}    \;.
\end{equation*}
After setting $\beta_N=\alpha_N$ \eqref{bigD-cDidentity1} becomes
\begin{equation}\label{bigD-cDidentity2}
\begin{split}
&Df_n^A-\cD f_{n+1}^A=\\
&\sum_{k_i\geq2}^{n}\sum_{N\geq2}\sum_{m=0}^{N-2}\alpha_N\,\binom{N}{m+1}\,(\iota_{k_1}...\iota_{k_m}DA_{k_{m+1}})\bullet(\iota_{k_{m+2}}...\iota_{k_{N-1}}\cD A_{k_N})\\
&+\sum_{k_i\geq2}^{n-1}\sum_{N\geq2}\alpha_N\sum_{m=0}^{N-2}\tfrac12\,\binom{N}{m+1}(\iota_{k_1}...\iota_{k_m}DA_{k_{m+1}})\bullet(\iota_{k_{m+2}}...\iota_{k_{N-1}}DA_{k_N})\\
&+\sum_{k_i\geq2}^{n+1}\sum_{N\geq2}\alpha_N\sum_{m=0}^{N-2}\tfrac12\,\binom{N}{m+1}(\iota_{k_1}...\iota_{k_m}\cD A_{k_{m+1}})\bullet(\iota_{k_{m+2}}...\iota_{k_{N-1}}\cD A_{k_N})\\
&-\sum_{k_i\geq2}^{n-1}\sum_{N\geq1}\alpha_N\,(-1)^{k_1+...+k_{N-1}}\iota_{k_1}...\iota_{k_{N-1}}\cL_{\Omega_2}A_{k_N}\;,
\end{split}    
\end{equation}
that has to be compared to the $\Omega$-independent part of the r.h.s. of \eqref{Bianchiproof1}:
\begin{equation}
\begin{split}
&\tfrac12\sum_{k=2}^{n-1}f_k^A\bullet f_{n+1-k}^A =\\
&\sum_{k_i\geq2}^{n}\sum_{N\geq2}\sum_{m=0}^{N-2}\alpha_{m+1}\alpha_{N-m-1}\,(\iota_{k_1}...\iota_{k_m}DA_{k_{m+1}})\bullet(\iota_{k_{m+2}}...\iota_{k_{N-1}}\cD A_{k_N})\\
&+\tfrac12\sum_{k_i\geq2}^{n-1}\sum_{N\geq2}\sum_{m=0}^{N-2}\alpha_{m+1}\alpha_{N-m-1}\,(\iota_{k_1}...\iota_{k_m}DA_{k_{m+1}})\bullet(\iota_{k_{m+2}}...\iota_{k_{N-1}}DA_{k_N})\\
&+\tfrac12\sum_{k_i\geq2}^{n+1}\sum_{N\geq2}\sum_{m=0}^{N-2}\alpha_{m+1}\alpha_{N-m-1}\,(\iota_{k_1}...\iota_{k_m}\cD A_{k_{m+1}})\bullet(\iota_{k_{m+2}}...\iota_{k_{N-1}}\cD A_{k_N})\;,
\end{split}  
\end{equation}
thus requiring $\alpha_N\,\binom{N}{m+1}=-\alpha_{m+1}\alpha_{N-m-1}\,$. By setting $c_N:=N!\,\alpha_N$ the requirement reads $c_N=-c_{m+1}c_{N-m-1}\,$, thus fixing $c_N=(-1)^{N-1}$ and
\begin{equation}
\alpha_N=\beta_N=\frac{(-1)^{N-1}}{N!}\;.    
\end{equation}
From \eqref{Bianchiproof1} one is left to prove
\begin{equation}\label{Bianchiproof2}
\begin{split}
&\cD f_{n+1}^\Omega-Df_n^\Omega+\sum_{k_i\geq2}^{n-1}\sum_{N\geq1}\alpha_N\,(-1)^{k_1+...+k_{N-1}}\iota_{k_1}...\iota_{k_{N-1}}\cL_{\Omega_2}A_{k_N}\\
&=\sum_{k=0}^{n-1}\Big[(\Omega_k+f^\Omega_k)\bullet f_{n+1-k}^A+(\Omega_k+\tfrac12\,f^\Omega_k)\bullet f_{n+1-k}^\Omega\Big]\;.
\end{split}
\end{equation}
The term $Df_n^\Omega$ is no different from $Df_n^A$ and obeys the same identity \eqref{Didentity}:
\begin{equation}\label{DidentityOmega}
\begin{split}
Df_n^\Omega&=\sum_{k_i\geq2}^n\sum_{N\geq2}\gamma_N\,D[i_{k_1}...i_{k_{N-1}}\Omega_{k_N}]\\
&=\sum_{k_i\geq2}^n\sum_{N\geq2}\gamma_N\sum_{m=0}^{N-2}\binom{N-1}{m+1}\,(\iota_{k_1}...\iota_{k_m}DA_{k_{m+1}})\bullet(\iota_{k_{m+2}}...\iota_{k_{N-1}}\Omega_{k_N})\\
&+\sum_{k_i\geq2}^n\sum_{N\geq2}\gamma_N\,(-1)^{k_1+...+k_{N-1}}\iota_{k_1}...\iota_{k_{N-1}}D\Omega_{k_N}\;,
\end{split}
\end{equation}
while one has to be more careful with $\cD f_{n+1}^\Omega\,$, since the identity \eqref{cDidentity} does not apply immediately, given that $\cD\Omega_2$ does not exist. We have instead
\begin{equation}
\begin{split}
-\cD f_{n+1}^\Omega&=-\sum_{k_i\geq2}^{n+1}\sum_{N\geq2}\gamma_N\,\cD[\iota_{k_1}...\iota_{k_{N-1}}\Omega_{k_N}]\\
&=\sum_{k_i\geq2}^{n+1}\sum_{N\geq2}\gamma_N\,\Big\{\sum_{l=0}^{N-3}(-1)^{k_1+...+k_l}\iota_{k_1}...\iota_{k_{l}}\big[\cD A_{k_{l+1}}\bullet(\iota_{k_{l+2}}...\iota_{k_{N-1}}\Omega_{k_N})\big]\\
&\hspace{28mm}-(-1)^{k_1+...+k_{N-2}}\iota_{k_1}...\iota_{k_{N-2}}\cD(A_{k_{N-1}}\bullet\Omega_{k_N})\Big\}  \;.  
\end{split}
\end{equation}
At this point one has to treat the ${k_N=2}$ term in the sum separately. Using in particular $\cD(A_{k_{N-1}}\bullet\Omega_2)=\cL_{\Omega_2}A_{k_{N-1}}-\cD A_{k_{N-1}}\bullet\Omega_2$ we get
\begin{equation}
\begin{split}
&\cD f_{n+1}^\Omega=-\sum_{k_i\geq2}^{n+1}\sum_{N\geq2}\gamma_N\sum_{m=0}^{N-2}\binom{N-1}{m+1}(\iota_{k_1}...\iota_{k_{m}}\cD A_{k_{m+1}})\bullet(\iota_{k_{m+2}}...\iota_{k_{N-1}}\Omega_{k_N})\\
&+\sum_{k_i\geq2}^n\sum_{N\geq2}\gamma_N\,(-1)^{k_1+...+k_{N-1}}\iota_{k_1}...\iota_{k_{N-1}}\cD\Omega_{k_N+1}+\sum_{k_i\geq2}^{n-1}\sum_{N\geq1}\gamma_{N+1}\,(-1)^{k_1+...+k_{N-1}}\iota_{k_1}...\iota_{k_{N-1}}\,\cL_{\Omega_2}A_{k_{N}}\;.
\end{split}    
\end{equation}
This finally yields
\begin{equation}\label{D-cDOmega}
\begin{split}
Df_n^\Omega-\cD f_{n+1}^\Omega&=\sum_{k_i\geq2}^{n+1}\sum_{N\geq2}\gamma_N\sum_{m=0}^{N-2}\binom{N-1}{m+1}\big[\iota_{k_1}...\iota_{k_{m}}(DA_{k_{m+1}-1}+\cD A_{k_{m+1}})\big]\bullet(\iota_{k_{m+2}}...\iota_{k_{N-1}}\Omega_{k_N})\\
&+\sum_{k_i\geq2}^n\sum_{N\geq2}\gamma_N\,(-1)^{k_1+...+k_{N-1}}\iota_{k_1}...\iota_{k_{N-1}}(D\Omega_{k_N}-\cD\Omega_{k_N+1})\\
&-\sum_{k_i\geq2}^{n-1}\sum_{N\geq1}\gamma_{N+1}\,(-1)^{k_1+...+k_{N-1}}\iota_{k_1}...\iota_{k_{N-1}}\,\cL_{\Omega_2}A_{k_{N}}    
\end{split}
\end{equation}
and, demanding that the $\cL_{\Omega_2}$ terms cancel, fixes
\begin{equation}
\gamma_N=-\alpha_{N-1}=\frac{(-1)^{N-1}}{(N-1)!}\;.
\end{equation}
With this value of $\gamma_N$ it is easy to see that the first line of \eqref{D-cDOmega} matches the $(\Omega+f^\Omega)\bullet f^A$ term in \eqref{Bianchiproof2}, while the second line is rewritten
by using the identity \eqref{CSrelations}:
\begin{equation}
\begin{split}
&\sum_{k_i\geq2}^n\sum_{N\geq2}\gamma_N\,(-1)^{k_1+...+k_{N-1}}\iota_{k_1}...\iota_{k_{N-1}}(\cD\Omega_{k_N+1}-D\Omega_{k_N})\\
&=\tfrac12\sum_{k_i\geq2}^{n+1}\sum_{N\geq3}\gamma_{N-1}\,(-1)^{k_1+...+k_{N-2}}\iota_{k_1}...\iota_{k_{N-2}}(\Omega_{k_{N-1}}\bullet\Omega_{k_{N}})\\
&=\tfrac12\sum_{k_i\geq2}^{n+1}\sum_{N\geq3}\gamma_{N-1}\sum_{l=0}^{N-2}\binom{N-2}{l}\,(\iota_{k_1}...\iota_{k_l}\Omega_{k_{l+1}})\bullet(\iota_{k_{l+2}}...\iota_{k_{N-1}}\Omega_{k_N})\;,
\end{split}    
\end{equation}
where we used \eqref{Pascaltriangle} in the last step. With $\gamma_N=\frac{(-1)^{N-1}}{(N-1)!}$ it is again easy to see that the above expression takes care of the final $(\Omega+\tfrac12\,f^\Omega)\bullet f^\Omega$ term of \eqref{Bianchiproof2}, finishing the proof of the Bianchi identity \eqref{Bianchiallorders}. 
\subsection{Gauge covariance}

Here our goal is to use the Bianchi identities 
\begin{equation}\label{Bianchirepeated}
D\cF_n+\tfrac12\sum_{k=2}^{n-1}\cF_k\bullet\cF_{n+1-k}=\cD\cF_{n+1}    
\end{equation}
to prove gauge covariance of the curvatures by induction. It is only at this point that we demand the differential $\cD$ to have trivial cohomology. 
Although it is necessary for this indirect proof, explicit computations for low degrees show that almost certainly this assumption is not actually needed, 
but we leave the necessary explicit formulation of the gauge transformations to future work. 
 
As before, the lowest case $n=2$ is proven directly, which we briefly recall in the present notation: One starts from
\begin{equation}
F_2=dA_1-\tfrac12\,A_1\circ A_1\;,    
\end{equation}
and postulates the gauge symmetry
\begin{equation}\label{deltaA1}
\delta A_1=D\lambda_0-\cD\lambda_1  \;.  
\end{equation}
The general variation of $F_2$ is given by
\begin{equation}
\delta F_2=d\delta A_1-\tfrac12\,(\delta A_1\circ A_1+A_1\circ\delta A_1)   =D\delta A_1+\tfrac12\,\cD(A_1\bullet\delta A_1) \;,
\end{equation}
that, upon using \eqref{deltaA1}, yields
\begin{equation}
\begin{split}
\delta F_2&=-F_2\circ\lambda_0+\cD\Big(\tfrac12\,A_1\bullet(D\lambda_0-\cD\lambda_1)-D\lambda_1\Big)\\
&=\cL_{\lambda_0}F_2+\cD\Big(\tfrac12\,A_1\bullet(D\lambda_0-\cD\lambda_1)-D\lambda_1-F_2\bullet\lambda_0\Big)\\
&=\cL_{\lambda_0}F_2+\cD\Delta_2\;.
\end{split}   
\end{equation}
One now introduces the two-form gauge field $A_2$ and the full two-form curvature as $\cF_2=F_2+\cD A_2\,$, obeying
\begin{equation}
\begin{split}
\delta\cF_2&=\delta(F_2+\cD A_2)=\cL_{\lambda_0}F_2+\cD(\delta A_2+\Delta_2)\\
&=\cL_{\lambda_0}\cF_2+\cD(\delta A_2-\cL_{\lambda_0}A_2+\Delta_2)
\end{split}    
\end{equation}
so that, by adjusting
\begin{equation}
\delta A_2=\cL_{\lambda_0}A_2-\Delta_2 -\cD\tilde\lambda_2   
\end{equation}
it is possible to achieve $\delta\cF_2=\cL_{\lambda_0}\cF_2\,$. Notice that, by using the explicit form of $\Delta_2$ we get the usual transformation
\begin{equation}
\Delta A_2:=\delta A_2+\tfrac12\,A_1\bullet\delta A_1=D\lambda_1+\cF_2\bullet\lambda_0-\cD\lambda_2  \;,
\end{equation}
with $\lambda_2=\tilde\lambda_2-\lambda_0\bullet A_2\,$.
Suppose next that we have fixed the gauge transformations of gauge fields $A_p$ with $1\leq p\leq n$ so that 
\begin{equation}
\delta\cF_p=\cL_{\lambda_0}\cF_p\;,\quad  \text{for}\quad 2\leq p\leq n  \;.
\end{equation}
We split $\cF_{n+1}=F_{n+1}+\cD A_{n+1}\,$, since $F_{n+1}$ only contains gauge fields $A_p$ with $p\leq n\,$. The Bianchi identity \eqref{Bianchirepeated}, thanks to covariance and Leibniz property of the Lie derivative, ensures that
\begin{equation}
\cD(\delta F_{n+1}-\cL_{\lambda_0}F_{n+1})=0\;.    
\end{equation}
Assuming that the differential $\cD$ has trivial cohomology, we can thus write
\begin{equation}
\delta F_{n+1}=\cL_{\lambda_0}F_{n+1}+\cD\Delta_{n+1}\;, 
\end{equation}
and
\begin{equation}
\begin{split}
\delta\cF_{n+1}&=\cL_{\lambda_0}F_{n+1}+\cD\Delta_{n+1}+\cD\delta A_{n+1}\\ &=\cL_{\lambda_0}\cF_{n+1}+\cD\big(\delta A_{n+1}-\cL_{\lambda_0}A_{n+1}+\Delta_{n+1}\big) \;, 
\end{split}
\end{equation}
such that, by fixing
\begin{equation}
\delta A_{n+1}=\cL_{\lambda_0}A_{n+1}-\Delta_{n+1}-\cD\lambda_{n+1}\;,
\end{equation}
the covariance of $\cF_{n+1}$ is established, and hence proven for all $n\,$.
\\
As a final comment, we notice that by taking the formal sum of the higher form potentials, as well as the pseudo CS forms and curvatures:
\begin{equation}
\cA:=\sum_{p=2}^\infty A_p\;,\quad \mathit{\Omega}:=\sum_{p=2}^\infty\Omega_p\;,\quad \cF:=\sum_{p=2}^\infty\cF_p\;,  
\end{equation}
it is possible to recast the Bianchi identities, as well as the definition of the curvatures, in the compact form
\begin{equation}
\begin{split}
&D\cF+\tfrac12\,\cF\bullet\cF=\cD\cF  \;,\\
&\cF=\sum_{N=0}^\infty\frac{(-\iota_{\cA})^N}{N!}\Big[\tfrac{1}{N+1}\,(D+\cD)\cA+\mathit{\Omega}\Big]\;.
\end{split}    
\end{equation}

In this paper we have discussed the algebraic structure underlying general tensor hierarchies purely at the kinematical level, determining gauge covariant curvatures of arbitrary rank together with their Bianchi identities. In order to provide dynamics to the above construction,\footnote{\emph{Note added in proof:} during the review stage, dynamical field equations were provided in terms of first-order duality relations in \cite{Bonezzi:2019bek}, where possible action principles have also been discussed.} natural candidates are action principles (or pseudo-actions supported by first-order duality relations) of the schematic form $\int\langle\cF\wedge,\star\cM\cF\rangle\,$, where $\langle\cdot,\cdot\rangle$ denotes a suitable inner product and $\cM$ plays the role of generalized metric built out of the scalar fields of the theory.

\section{Topological theories based on $L_{\infty}$ algebras}

Our goal in this section is to relate the infinity enhanced Leibniz algebra introduced above to the closely related $L_{\infty}$ algebras. 
It is known that for an `enhanced Leibniz algebra' as defined in \cite{Strobl:2019hha}, consisting of the vector space $X_0\oplus X_1\,$, there is an 
associated Lie 2-algebra, \emph{i.e.}~an $L_\infty$ algebra whose highest bracket is the three bracket $l_3\,$. Similarly, given the structure of an infinity enhanced Leibniz algebra, one can define $L_\infty$ algebras characterized by a nilpotent operator $l_1\equiv\cD\,$, and graded antisymmetric brackets $l_n$ obeying higher Jacobi-like relations. However, we will show that, in contrast to Leibniz algebras and extensions thereof, the $L_\infty$ structure alone is not sufficient to define gauge covariant curvatures for higher form potentials. It is still possible, by means of $L_\infty$ brackets alone, to define field strengths that transform into themselves under gauge transformations. This allows us to define topological tensor hierarchies, whose field equations amount to zero curvature conditions.

\subsection{Warm-up}

We begin with a warm-up example, trying to construct curvatures for the lowest gauge potentials by means of $L_\infty$ brackets. This will show problems already at the level of the three-form curvature, essentially due to the lack of a proper covariant derivative. We will then proceed to construct a topological tensor hierarchy to all orders in terms of a general $L_\infty$ algebra, not necessarily based on an underlying Leibniz algebra.
\\
We consider here a set of differential forms and gauge parameters taking value in an $L_\infty$ algebra, with $L_\infty$ degree assignments $|A_\mu|=|\lambda|=0\,$, $|A_{\mu\nu}|=|\lambda_\mu|=1$ and so on.
More specifically, we identify 
 \be
  \begin{split}
   \text{`trivial parameters':}& \qquad \chi_0\;, \; \chi_1\;, \ldots \\
   \text{gauge parameters:}& \qquad \lambda_0\;, \; \lambda_1\;, \; \lambda_2\;, \ldots\\
   \text{gauge fields:}&\qquad  A_1\;, \; A_2\;, \; A_3\;, \ldots\\
   \text{field strengths:}&\qquad  {\cal F}_2\;, \; {\cal F}_3\;,  \ldots\;, 
  \end{split}
 \ee  
where now the index denotes the form degree and the $L_{\infty}$ degree can be inferred 
from the following diagram:  
\be\label{DIAGRAM}
\begin{array}{ccccccccccc}\cdots&\xlongrightarrow{l_1} &X_{[0]}^2 {}  &\xlongrightarrow{l_1}
&X_{[0]}^1 (\chi_0)&
\xlongrightarrow{l_1}&X^0_{[0]} ( \lambda_0)
\\[1.5ex]
\Big{\downarrow}{{\rm d}}&&\Big{\downarrow}{{\rm d}}&&\Big{\downarrow}{{\rm d}}&&\Big{\downarrow}{{\rm d}}
\\[1.5ex]
\cdots&\xlongrightarrow{l_1}&X^2_{[1]}(\chi_1) &\xlongrightarrow{l_1}&X^1_{[1]}(\lambda_1)&
\xlongrightarrow{l_1}&X^0_{[1]}(A_1)\\[1.5ex]
\Big{\downarrow}{{\rm d}}&&\Big{\downarrow}{{\rm d}}&&\Big{\downarrow}{{\rm d}}&&\Big{\downarrow}{{\rm d}}
\\[1.5ex]
\cdots&\xlongrightarrow{l_1}&X^2_{[2]}(\lambda_{2}) &\xlongrightarrow{l_1}&X^1_{[2]}(A_{2})&
\xlongrightarrow{l_1}&X^0_{[2]}({\cal F}_{2})\\[1.5ex]
\Big{\downarrow}{{\rm d}}&&\Big{\downarrow}{{\rm d}}&&\Big{\downarrow}{{\rm d}}&&\Big{\downarrow}{{\rm d}}
\\[1.5ex]
\cdots&\xlongrightarrow{l_1}&X^2_{[3]}(A_3) &\xlongrightarrow{l_1}&X^1_{[3]} ({\cal F}_3)&
\xlongrightarrow{l_1}&X^0_{[3]} ({\rm d}{\cal F}_2)
\end{array}
\ee
Note that the sequence of gauge parameters, fields, field strengths, respectively, 
runs `diagonally' through the diagram (from north-east to south-west), 
in principle indefinitely. 

We start again, in parallel with the construction of section 2,  by postulating the gauge symmetry for the one-form as
\begin{equation}
\delta A_\mu=\del_\mu\lambda-l_2(A_\mu,\lambda)-l_1\lambda_\mu \;.  
\end{equation}
Here $l_1$ is the nilpotent, degree $-1\,$, operator of the $L_\infty$ algebra, that is a differential w.r.t. the $l_2$ bracket:
\begin{equation}\label{l1l2}
l_1 l_2(a,b)=l_2(l_1a,b)+(-1)^{|a|}l_2(a,l_1b)   \;, 
\end{equation}
and can be identified with the $\cD$ operator of the Leibniz algebra of the previous sections.
The only other $L_\infty$ relation needed at this level is
\begin{equation}\label{l1l3}
\begin{split}
0 &= l_1\,l_3(a,b,c)+l_3(l_1a,b,c)+(-1)^{|a|}l_3(a,l_1b,c)+(-1)^{|a|+|b|}l_3(a,b,l_1c)\\
&+l_2(l_2(a,b),c)+(-1)^{|a|(|b|+|c|)}l_2(l_2(b,c),a)+(-1)^{|c|(|a|+|b|)}l_2(l_2(c,a),b)\;,
\end{split}    
\end{equation}
recalling that $l_1$ does not act on degree zero elements.
By using \eqref{l1l2} and \eqref{l1l3} one still finds that the two-form curvature defined by 
 \begin{equation}
\cF_{\mu\nu}=2\,\del_{[\mu} A_{\nu]}-l_2(A_\mu,A_\nu)+l_1\, B_{\mu\nu}     
 \end{equation}
transforms covariantly: $\delta\cF_{\mu\nu}=l_2(\lambda,\cF_{\mu\nu})\,$, provided that the two-form gauge field transforms as
\begin{equation}
\delta B_{\mu\nu}=2\,\del_{[\mu}\lambda_{\nu]}-2\,l_2(A_{[\mu},\lambda_{\nu]})+l_2(\lambda,B_{\mu\nu})-l_3(A_\mu,A_\nu,\lambda)-l_1\,\lambda_{\mu\nu} \;.   
\end{equation}
This apparently suggests to define a ``covariant'' derivative: $D_\mu x:=\del_\mu x-l_2(A_\mu,x)$ that allows to rewrite
\begin{equation}\label{deltaAB}
\begin{split}
&\delta A_\mu=D_\mu\lambda-l_1\lambda_\mu\;,\\
&\delta B_{\mu\nu}=2\,D_{[\mu}\lambda_{\nu]}+l_2(\lambda,B_{\mu\nu})-l_3(A_\mu,A_\nu,\lambda)-l_1\,\lambda_{\mu\nu} 
\;,
\end{split}    
\end{equation}
but is effectively of little use, since $D_\mu$ is \emph{not} a covariant operation in the usual sense. Indeed,
given any element $x$
that transforms covariantly in the sense that 
\begin{equation}
 \delta x=l_2(\lambda,x)\;,
 \end{equation}
one has
\begin{equation}\label{covariancefailure}
 \delta\,D_\mu x=l_2(\lambda,D_\mu x)+l_1 l_3(A_\mu,x,\lambda)+l_3(A_\mu,l_1 x,\lambda)\;, 
 \end{equation}
which contains bare gauge fields.  
At this stage, the most general ansatz for the three-form curvature is
\begin{equation}
\cH_{\mu\nu\lambda}=3\,\del_{[\mu} B_{\nu\lambda]}+\alpha\,l_2(A_{[\mu},B_{\nu\lambda]})+\beta\,l_3(A_\mu,A_\nu,A_\lambda)+l_1\, C_{\mu\nu\lambda}  \;,  
\end{equation}
 and using \eqref{deltaAB} one finds the variation
\begin{equation}
\begin{split}
\delta\cH_{\mu\nu\rho} &= (\alpha+3)\,l_2(\del_{[\mu}\lambda,B_{\nu\rho]})+3(\beta-1)\,l_3(\del_{[\mu}\lambda,A_\nu,A_{\rho]})+2(\alpha+3)\,l_2(A_{[\mu},\del_\nu\lambda_{\rho]})+...\\
&+l_1\big(\delta C_{\mu\nu\rho}-3\,\del_{[\mu}\lambda_{\nu\rho]}+...\big)\;,
\end{split}
\end{equation}
where the omitted terms do not contain derivatives of the gauge parameters. This already fixes $\alpha=-3$ and $\beta=1\,$, yielding
\begin{equation}
\cH_{\mu\nu\lambda}=3\,D_{[\mu} B_{\nu\lambda]}+l_3(A_\mu,A_\nu,A_\lambda)+l_1\, C_{\mu\nu\lambda}    \;.
\end{equation}
The $l_1$-exact terms in $\delta\cH_{\mu\nu\rho}$ can be absorbed by setting\footnote{To find this one also needs the relation $l_1\,l_4(x_1,x_2,x_3,x_4)=4\,l_2(l_3(x_{[1},x_2,x_3),x_{4]})-6\,l_3(l_2(x_{[1},x_2),x_3,x_{4]})$ for $|x_i|=0\,$.}
\begin{equation}
\begin{split}
\delta C_{\mu\nu\rho}\ = \ &\,3\,D_{[\mu}\lambda_{\nu\rho]}+l_2(\lambda,C_{\mu\nu\rho})-3\,l_2(\lambda_{[\mu},B_{\nu\rho]})\\
&-3\,l_3(\lambda,A_{[\mu},B_{\nu\rho]})-3\,l_3(\lambda_{[\mu},A_\nu,A_{\rho]})+l_4(\lambda,A_\mu,A_\nu,A_\rho)    \;,
\end{split}
\end{equation} 
but one is still left with
\begin{equation}\label{deltaH}
\delta\cH_{\mu\nu\rho}=l_2(\lambda,\cH_{\mu\nu\rho})+3\,l_2(\lambda_{[\mu},\cF_{\nu\rho]})+3\,l_3(\lambda,A_{[\mu},\cF_{\nu\rho]})   \;,
\end{equation}
that is not covariant even in terms of a formal sum of curvatures, due to the bare gauge field in the last term. 
However, since $\delta\cH_{\mu\nu\rho}$ is proportional to itself and $\cF_{\mu\nu}\,$, it still allows for the zero curvature conditions
\begin{equation}
\cH_{\mu\nu\rho}=0\;,\quad \cF_{\mu\nu}=0    \;, 
\end{equation}
to be gauge invariant topological field equations.
\\
After this explicit example for the lowest ranks of the tensor hierarchy, we will now turn to determine the gauge transformations and field strengths for differential forms $A_{\mu_1...\mu_p}$ of arbitrary degree taking values in an $L_\infty$ algebra, that are consistent as topological field equations of the form $\cF_{\mu_1...\mu_{p+1}}=0\,$.

\subsection{Differential forms taking values in $L_{\infty}$ algebras}

In this section we will only assume that the field content consists of a set of differential forms of arbitrary degree, taking values in an $L_\infty$ algebra with multilinear, graded antisymmetric, maps $l_n$ of intrinsic degree $n-2$, obeying the quadratic relations
\begin{equation}
\sum_{n=0}^{N-1}(-1)^{n(N+1)}l_{n+1}l_{N-n}=0\;,\quad N=1,2,...,\infty \;,   
\end{equation}
where $N$ is the total number of arguments involved. The left hand side is meant to act on the graded antisymmetrized tensor algebra so that, for instance, the $N=3$ relation ${l_1l_3+l_2l_2+l_3l_1=0}$ explicitly reads as \eqref{l1l3}. The set of differential forms organizes naturally in terms of a double grading $(p,d)$ given by the form degree and the $L_\infty$ degree, respectively. A generic differential $p$-form of $L_\infty$ degree $d$ will be denoted by
\begin{equation}
\omega_{\mu_1...\mu_p}^d \ \equiv \ \omega_{\mu[p]}^d\;.    
\end{equation}
 This two-dimensional array has the structure of a bi-complex with respect to two separate differentials, the de Rham differential $d\,$, increasing the form degree by one, and the $l_1$ operator, decreasing the $L_\infty$ degree by one. The two gradings, in spite of being independent, can be linked in the tensor hierarchy thanks to the physical interpretation of the fields: A $p$-form curvature has $L_\infty$ degree ${p-2}\,$, $F_{\mu[p]}^{p-2}\,$, a $p$-form gauge field 
 $A_{\mu[p]}^{p-1}$ has degree\footnote{This is chosen to give $L_\infty$ degree zero to the vector field, its scalar gauge parameter and curvature two-form.} $p-1\,$, a $p$-form gauge parameter has degree $p\,$, $\xi_{\mu[p]}^p$ and so on. One can notice that a new, diagonal, degree can be defined as the difference between $L_\infty$ and form degree: $N:=d-p\,$, and that this degree is only sensitive to the physical role of a given differential form. Indeed, it is immediate to see that every curvature $F_{\mu[p]}^{p-2}$ has $N$-degree $-2\,$, every gauge field $A_{\mu[p]}^{p-1}$ has $N$-degree $-1$ and any gauge parameter $\xi_{\mu[p]}^p$ has $N$-degree $0\,$. The main advantage of classifying the fields in terms of the $N$-degree (that is identified with their physical interpretation) is that gauge fields (as well as curvatures, gauge parameters etc.) of arbitrary form degree can be treated on equal footing, and eventually dealt with at once in a single, string field-like, object. In the following, we will show that it is possible to define new maps, that we will denote by $\ell_n\,$, that obey the usual $L_\infty$ symmetry properties and quadratic relations in terms of the single degree $N\,$. The new maps differ from the $l_n$'s only by phases that are quite lengthy to determine. In order to deal efficiently with such phases we introduce anticommuting generating elements $\theta^\mu\,$, obeying
\begin{equation}
\theta^\mu\theta^\nu+\theta^\nu\theta^\mu =0\;.   
\end{equation}
They differ from ordinary one-form basis elements $dx^\mu$ in that they are formally assigned $L_\infty$ degree $-1\,$. By this we mean that they pick up a sign  when commuting with a differential form: $\theta^\mu\,\omega_{\nu[p]}^d=(-1)^d\omega_{\nu[p]}^d\,\theta^\mu\,$, where $d$ is the $L_{\infty}$ degree, 
as well as when (formally) going through an $l_n$ map: $\theta^\mu\,l_n=(-1)^nl_n\,\theta^\mu\,$. This allows us to define objects
\begin{equation}
\omega_p^{N}:=\tfrac{1}{p!}\,\theta^{\mu_1}...\theta^{\mu_p}\,\omega_{\mu_1...\mu_p}^d\equiv \theta^{\mu[p]}\, \omega_{\mu[p]}^d    
\end{equation}
that have (so far formally) the $N$-degree as 
their $L_\infty$ degree. The precise way to determine the signs  defining the new $\ell_n$ maps is to move all the $\theta$ generating elements to the left and outside the maps according to the above commutation relations, \emph{e.g.}
\begin{equation}
\begin{split}
&\ell_1(\omega_p^N)=\ell_1(\theta^{\mu[p]}\,\omega_{\mu[p]}^d):=(-1)^p\,\theta^{\mu[p]}\,l_1(\omega_{\mu[p]}^d)    \\
&\ell_2(\omega_{p_1}^{N_1},\omega_{p_2}^{N_2})=\ell_2\left(\theta^{\mu[p_1]}\,\omega_{\mu[p_1]}^{d_1},\,\theta^{\nu[p_2]}\,\omega_{\nu[p_2]}^{d_2}\right):=(-1)^{p_2d_1}\theta^{\mu[p_1]}\theta^{\nu[p_2]}\,l_2\left(\omega_{\mu[p_1]}^{d_1},\,\omega_{\nu[p_2]}^{d_2}\right)\;.
\end{split}
\end{equation}
This allows us to give a precise definition of the $\ell_n$ maps as
\begin{equation}\label{extendedells}
\begin{split}
&\ell_n(\omega_{p_1}^{N_1},...,\omega_{p_n}^{N_n}):=(-1)^\Sigma\,\theta^{\mu[p_1]}...\theta^{\nu[p_n]}\,l_n\left(\omega_{\mu[p_1]}^{d_1},...,\omega_{\mu[p_n]}^{d_n}\right)  \;,\\
&\Sigma=n\sum_{i=1}^np_i+\sum_{i=2}^n\sum_{j=1}^{i-1}p_i\,d_j\;.
\end{split}    
\end{equation}
According to this definition, the $\ell_n$ maps are graded antisymmetric w.r.t. the $N$-degree, as can be proven by direct computation:
\begin{equation}
\begin{split}
&\ell_n(\omega_{p_1}^{N_1},...,\omega_{p_i}^{N_i},\omega_{p_{i+1}}^{N_{i+1}},...,\omega_{p_n}^{N_n}) =(-1)^{n\sum_{k=1}^np_k+\sum_{k=2}^n\sum_{l=1}^{k-1}p_k\,d_l}\theta^{\mu[p_1]}...\theta^{\nu[p_i]}\theta^{\lambda[p_{i+1}]}...\theta^{\rho[p_n]}\\
&\hspace{50mm}\;\times l_n\left(\omega_{\mu[p_1]}^{d_1},...,\omega_{\nu[p_i]}^{d_i},\omega_{\lambda[p_{i+1}]}^{d_{i+1}},...,\omega_{\rho[p_n]}^{d_n}\right)\\
&=(-1)^{n\sum_{k=1}^np_k+\sum_{k=2}^n\sum_{l=1}^{k-1}p_k\,d_l}(-1)^{1+p_ip_{i+1}+d_id_{i+1}}\theta^{\mu[p_1]}...\theta^{\lambda[p_{i+1}]}\theta^{\nu[p_i]}...\theta^{\rho[p_n]}\\
&\hspace{50mm}\;\times l_n\left(\omega_{\mu[p_1]}^{d_1},...,\omega_{\lambda[p_{i+1}]}^{d_{i+1}},\omega_{\nu[p_i]}^{d_i},...,\omega_{\rho[p_n]}^{d_n}\right)\\
&=(-1)^{1+N_iN_{i+1}}\ell_n(\omega_{p_1}^{N_1},...,\omega_{p_{i+1}}^{N_{i+1}},\omega_{p_i}^{N_i},...,\omega_{p_n}^{N_n})\;.
\end{split}    
\end{equation}
The $n$-dependent part of the sign factor in the definition \eqref{extendedells} (that corresponds to the formal property of $\theta$'s picking a phase to go through the maps themselves) is needed in order to ensure the correct symmetry property of nested $\ell_n$ maps, for instance
\begin{equation}
\ell_2(\ell_n(\omega_{p_1}^{N_1},...,\omega_{p_n}^{N_n}),\eta_q^{N_q})=(-1)^{1+N_q(\sum_iN_i+n-2)}\,\ell_2(\eta_q^{N_q},\ell_n(\omega_{p_1}^{N_1},...,\omega_{p_n}^{N_n}))   \;,
\end{equation}
and in general
\begin{equation}
\begin{split}
&\ell_n(\omega_{p_1}^{N_1},...,\ell_m(\eta_{q_1}^{M_1},...,\eta_{q_m}^{M_m}),\omega_{p_k}^{N_k},...,\omega_{p_{n-1}}^{N_{n-1}})\\
&=(-1)^{1+N_k(\sum_iM_i+m-2)}\, \ell_n(\omega_{p_1}^{N_1},...,\omega_{p_k}^{N_k},\ell_m(\eta_{q_1}^{M_1},...,\eta_{q_m}^{M_m}),...,\omega_{p_{n-1}}^{N_{n-1}})   \;,
\end{split}    
\end{equation}
thereby confirming that the $\ell_n$ maps have intrinsic degree $n-2\,$. In order to prove that the new maps $\ell_n$ obey the same quadratic relations as the $l_n\,$, but w.r.t.~the $N$-degree, we have to show that they pick the correct sign under the permutation of arguments between the two maps involved: The original maps obey the quadratic relations
\begin{equation}
\begin{split}
&l_n(l_m(\omega_{\mu[p_1]}^{d_1},...,\omega_{\mu[p_m]}^{d_m}),\omega_{\mu[p_{m+1}]}^{d_{m+1}},...,\omega_{\mu[p_{n+m-1}]}^{d_{n+m-1}})\\
&+(-1)^{1+d_md_{m+1}}l_n(l_m(\omega_{\mu[p_1]}^{d_1},...,\omega_{\mu[p_{m+1}]}^{d_{m+1}}),\omega_{\mu[p_{m}]}^{d_{m}},...,\omega_{\mu[p_{n+m-1}]}^{d_{n+m-1}})+...=0\;.
\end{split}
\end{equation}
Multiplying this relation by
$$
(-1)^{(n+m)\sum_{i=1}^{n+m-1}p_i+\sum_{i=2}^{n+m-1}\sum_{j=1}^{i-1}p_id_j}\,\theta^{\mu[p_1]}...\theta^{\mu[p_{n+m-1}]}
$$
one obtains
\begin{equation}
\begin{split}
&\ell_n(\ell_m(\omega_{p_1}^{N_1},...,\omega_{p_m}^{N_m}),\omega_{p_{m+1}}^{N_{m+1}},...,\omega_{p_{n+m-1}}^{N_{n+m-1}})\\
&+(-1)^{1+N_mN_{m+1}}\ell_n(\ell_m(\omega_{p_1}^{N_1},...,\omega_{p_{m+1}}^{N_{m+1}}),\omega_{p_{m}}^{N_{m}},...,\omega_{p_{n+m-1}}^{N_{n+m-1}})+...=0\;,
\end{split}
\end{equation}
thus proving that the $\ell_n$ maps obey the $L_\infty$ relations with respect to the $N$-degree.
The de Rham differential is now written as
\begin{equation}
d:=\theta^\mu\del_\mu    
\end{equation}
and thus possesses $N$-degree $-1\,$. Accordingly, it obeys a graded Leibniz rule that only sees the $N$-degree:
\begin{equation}
\begin{split}
d\,\ell_n(\omega_{p_1}^{N_1},\omega_{p_2}^{N_2},...,\omega_{p_n}^{N_n})&=  (-1)^n\big\{\ell_n(d\omega_{p_1}^{N_1},\omega_{p_2}^{N_2},...,\omega_{p_n}^{N_n})+(-1)^{N_1}\ell_n(\omega_{p_1}^{N_1},d\omega_{p_2}^{N_2},...,\omega_{p_n}^{N_n})+\\
&...+(-1)^{\sum_{i=1}^{n-1}N_i}\ell_n(\omega_{p_1}^{N_1},\omega_{p_2}^{N_2},...,d\omega_{p_n}^{N_n})\big\}\;.
\end{split}
\end{equation}
In particular, it obeys $d\ell_1+\ell_1d=0\,$, that allows one to define a new nilpotent operator:
\begin{equation}
\tilde\ell_1:=\ell_1+d    
\end{equation}
that still obeys the $L_\infty$ relations.

\subsection{Topological gauge theory }
We are now ready to construct a topological higher gauge theory with the above ingredients: The gauge fields $A_{\mu[p]}$ can be packaged in a single string field-like object of $N$-degree $-1$
\begin{equation}
\cA(x,\theta):=\sum_{p=1}^\infty\tfrac{1}{p!}\,\theta^{\mu_1}...\theta^{\mu_p}\,A_{\mu_1...\mu_p}(x)    \;,\quad N_{\cA}=-1\;, 
\end{equation}
and similarly one can define a degree zero gauge parameter
\begin{equation}
\Xi(x,\theta):=\sum_{p=0}^\infty\tfrac{1}{p!}\,\theta^{\mu_1}...\theta^{\mu_p}\,\xi_{\mu_1...\mu_p}(x)    \;,\quad N_{\Xi}=0 \;,   \end{equation}
and degree $-2$ curvature
\begin{equation}
\cF(x,\theta):=\sum_{p=2}^\infty\tfrac{1}{p!}\,\theta^{\mu_1}...\theta^{\mu_p}\,F_{\mu_1...\mu_p}(x)    \;,\quad N_{\cF}=-2 \;. \end{equation}
For the field strength and gauge transformations, we make the ansatz
\begin{equation}
\begin{split}
\delta\cA&=d\Xi+\sum_{n=1}^\infty\alpha_n\,\ell_n(\cA,...,\cA,\Xi)\\
\cF&=d\cA+\sum_{n=1}^\infty\gamma_n\,\ell_n(\cA,...,\cA)\;,
\end{split}
\end{equation}
where the coefficients $\alpha_n$ and $\gamma_n$ have to be fixed by demanding that the curvatures transform into themselves. For the gauge variation of the curvature we compute
\begin{equation}
\begin{split}
\delta\cF&=\sum_{n=1}^\infty(-1)^n(n-1)\alpha_n\,\ell_n(d\cA,\cA,...,\cA,\Xi)-\alpha_n\,\ell_n(\cA,...,\cA,d\Xi)\\
&+\sum_{n=1}^\infty n\gamma_n\,\Big\{\ell_n(d\Xi,\cA,...,\cA)+\sum_{m=1}^\infty\alpha_m\,\ell_n(\ell_m(\cA,...,\cA,\Xi),\cA,...,\cA)\Big\}\;,
\end{split}    
\end{equation}
and demanding that the $d\Xi$ terms cancel fixes $\gamma_n=\frac{1}{n}\alpha_n\,$, yielding
\begin{equation}\label{deltaF}
\begin{split}
\delta\cF&=\sum_{n=1}^\infty(-1)^n(n-1)\alpha_n\,\ell_n(d\cA,\cA,...,\cA,\Xi)+\sum_{n,m=1}^\infty \alpha_n\alpha_m\,\ell_n(\ell_m(\cA,...,\cA,\Xi),\cA,...,\cA)\\
&=\sum_{n=1}^\infty(-1)^n(n-1)\alpha_n\,\ell_n(\cF,\cA,...,\cA,\Xi)\\
&+\sum_{n,m=1}^\infty\alpha_n\alpha_m\Big\{\ell_n(\ell_m(\cA,...,\cA,\Xi),\cA,...,\cA)-(-1)^n\tfrac{n-1}{m}\,\ell_n(\ell_m(\cA,...,\cA),\cA,...,\cA,\Xi)\Big\}\;.
\end{split}    
\end{equation}
For the curvature to transform proportionally to itself the last line above has to vanish. The $L_\infty$ relations for $N-1$ $\cA$'s of degree $-1$ and one $\Xi$ of degree zero read
\begin{equation}
\begin{split}
\sum_{k=0}^{N-1}(-1)^{k(N+1)}&\Big[\tfrac{1}{(k-1)!(N-k)!}\,\ell_{k+1}(\ell_{N-k}(\cA,...,\cA),\cA,...,\cA,\Xi)\\
&+\tfrac{(-1)^k}{k!(N-1-k)!}\,\ell_{k+1}(\ell_{N-k}(\cA,...,\cA,\Xi),\cA,...,\cA)\Big]=0\;,    
\end{split}
\end{equation}
while the last line of \eqref{deltaF} can be rewritten as
\begin{equation}
\begin{split}
& \sum_{n,m=1}^\infty\alpha_n\alpha_m\Big\{\ell_n(\ell_m(\cA,...,\cA,\Xi),\cA,...,\cA)-(-1)^n\tfrac{n-1}{m}\,\ell_n(\ell_m(\cA,...,\cA),\cA,...,\cA,\Xi)\Big\}\\   
&=\sum_{N=1}^\infty\sum_{k=0}^{N-1}\alpha_{k+1}\alpha_{N-k}\Big\{\ell_{k+1}(\ell_{N-k}(\cA,...,\cA,\Xi),\cA,...,\cA)\\
&\hspace{40mm}+(-1)^k\tfrac{k}{N-k}\,\ell_{k+1}(\ell_{N-k}(\cA,...,\cA),\cA,...,\cA,\Xi)\Big\}\;.
\end{split}    
\end{equation}
For this to be proportional to the $L_\infty$ relations one has to demand that
\begin{equation}
\alpha_{k+1}\alpha_{N-k}=f(N)\frac{(-1)^{Nk}}{k!(N-1-k)!}   \end{equation}
for an arbitrary $f(N)\,$, which can be solved by $\alpha_n=\frac{(-1)^{\frac{n(n\pm1)}{2}}}{(n-1)!}\,$. By choosing the solution with the minus sign we get
\begin{equation}
\begin{split}
\delta\cA&=d\Xi+\sum_{n=1}^\infty\frac{(-1)^{\frac{n(n-1)}{2}}}{(n-1)!}\,\ell_n(\cA,...,\cA,\Xi)\;, \\
\cF&=d\cA+\sum_{n=1}^\infty\frac{(-1)^{\frac{n(n-1)}{2}}}{n!}\,\ell_n(\cA,...,\cA)\;, \end{split}    
\end{equation}
with the curvature transforming as
\begin{equation}\label{deltacFtop}
\delta\cF=\sum_{n=2}^\infty\frac{(-1)^{\frac{n(n+1)}{2}}}{(n-2)!}\,\ell_n(\cF,\cA,...,\cA,\Xi)    \;.
\end{equation}
As anticipated in the Introduction, the curvatures are not gauge covariant, due to the presence of bare gauge fields in \eqref{deltacFtop} that cannot be avoided. However, the zero curvature condition
\begin{equation}
\cF=0    
\end{equation}
is a consistent, gauge invariant, topological field equation.
The curvature obeys the generalized Bianchi identity
\begin{equation}
d\cF+\sum_{n=1}^\infty\frac{(-1)^{\frac{n(n-1)}{2}}}{(n-1)!}\,\ell_n(\cA,...,\cA,\cF)\equiv0   \;, 
\end{equation}
that suggests to define a generalized covariant derivative as
\begin{equation}
Dx:=dx+\sum_{n=1}^\infty\frac{(-1)^{\frac{n(n-1)}{2}}}{(n-1)!}\,\ell_n(\cA,...,\cA,x)  \;, 
\end{equation}
such that the Bianchi identity and gauge transformation of $\cA$ take the form
\begin{equation}
\delta\cA=D\Xi\;,\quad D\cF\equiv0\;.    
\end{equation}
Establishing the form of trivial gauge parameters is facilitated by the identity
\begin{equation}\label{Dsquaretop}
D^2x=  \sum_{n=2}^\infty\frac{(-1)^{\frac{n(n+1)}{2}}}{(n-2)!}\,\ell_n(\cF,\cA,...,\cA,x)  \;,  
\end{equation}
that generalizes the usual Yang-Mills relation $D^2=F\,$. It can be proven by direct computation:
\begin{equation}
\begin{split}
D^2x&=  \sum_{n=2}^\infty\frac{(-1)^{\frac{n(n+1)}{2}}}{(n-2)!}\,\ell_n(d\cA,\cA,...,\cA,x)-\sum_{n=1}^\infty\frac{(-1)^{\frac{n(n-1)}{2}}}{(n-1)!}\,\ell_n(\cA,...,\cA,dx) \\
&+\sum_{n=1}^\infty\frac{(-1)^{\frac{n(n-1)}{2}}}{(n-1)!}\,\ell_n(\cA,...,\cA,dx)+\sum_{n,m=1}^\infty\frac{(-1)^{\frac{n(n-1)+m(m-1)}{2}}}{(n-1)!(m-1)!}\,\ell_n(\cA,...,\cA,\ell_m(\cA,...,\cA,x))\\
&= \sum_{n=2}^\infty\frac{(-1)^{\frac{n(n+1)}{2}}}{(n-2)!}\,\ell_n(\cF,\cA,...,\cA,x)\\
&-\sum_{n,m=1}^\infty\frac{(-1)^{\frac{n(n+1)+m(m-1)}{2}}(n-1)}{(n-1)!m!}\,\ell_n(\ell_m(\cA,...,\cA),\cA,...,\cA,x)\\
&+\sum_{n,m=1}^\infty\frac{(-1)^{\frac{n(n-1)+m(m-1)}{2}}(-1)^{|x|(n-1)}}{(n-1)!(m-1)!}\,\ell_n(\ell_m(\cA,...,\cA,x)\cA,...,\cA)\\
&=\sum_{n=2}^\infty\frac{(-1)^{\frac{n(n+1)}{2}}}{(n-2)!}\,\ell_n(\cF,\cA,...,\cA,x)+\sum_{N=1}^\infty(-1)^{\frac{N(N-1)}{2}}\\
&\times\sum_{k=0}^{N-1}\frac{(-1)^{k(N+1)}}{k!(N-k-1)!}\Big[\tfrac{k}{N-k}\,\ell_{k+1}(\ell_{N-k}(\cA,...,\cA),\cA,...,\cA,x)\\
&\hspace{40mm}+(-1)^{k(|x|+1)}\ell_{k+1}(\ell_{N-k}(\cA,...,\cA,x),\cA,...,\cA)\Big]\\
&=\sum_{n=2}^\infty\frac{(-1)^{\frac{n(n+1)}{2}}}{(n-2)!}\,\ell_n(\cF,\cA,...,\cA,x)\;,
\end{split}    
\end{equation}
where we denoted the degree of $x$ by $|x|$ and used the $L_\infty$ relations in the last line. 
\\
As promised, the property \eqref{Dsquaretop} immediately shows that a gauge parameter of the form $\Xi=D\Lambda$ generates transformations that are trivial on-shell:
\begin{equation}
\delta_{D\Lambda}\cA=D^2\Lambda= \sum_{n=2}^\infty\frac{(-1)^{\frac{n(n+1)}{2}}}{(n-2)!}\,\ell_n(\cF,\cA,...,\cA,\Lambda)\stackrel{\cF=0}{=}0  \;.  
\end{equation}

The above construction is closely related to the Alexandrov-Kontsevich-Schwarz-Zaboronsky (AKSZ) \cite{Alexandrov:1995kv} formalism to construct topological sigma models from graded source manifolds (for a review see \emph{e.g.} \cite{Ikeda:2012pv}).
In particular, the quantity
\begin{equation}
Q(\cA):=\sum_{n=1}^\infty\frac{(-1)^{\frac{n(n-1)}{2}}}{n!}\,\ell_n(\cA,...,\cA)    
\end{equation}
appearing in the field equation
\begin{equation}
\cF=d\cA+Q(\cA)=0\;,    
\end{equation}
is related to the power series expansion of the Q-structure of the target space, in AKSZ terms. The present construction, however, only produces the ghost-number zero classical fields of the theory. In fact, contrary to the usual AKSZ construction, the oscillators $\theta^\mu$ do not carry ghost number and all component fields of $\cA$ are classical gauge fields. One can indeed think about constructing the entire BV spectrum of the model, by further endowing the $\theta^\mu$ oscillators with ghost number, and enlarge the field content by promoting every component $A_p$ in $\cA=\sum_pA_p$ to an arbitrary function of $\theta^\mu\,$.


\section{Conclusions and Outlook}

In this paper we have developed the general gauge theory of Leibniz-Loday algebras.  
We introduced the structure of an `infinity enhanced Leibniz algebra' and proved that there is 
an associated tensor hierarchy of $p$-form gauge potentials that is consistent to arbitrary levels. 
Our proposal is that the `infinity enhanced Leibniz algebra' yields  the proper mathematical axiomatization of the 
notion of `tensor hierarchy' developed in theoretical physics.

There are numerous potential applications and further extensions of the general framework developed here, 
which we briefly list in the following: 
\begin{itemize}
\item Various examples of Leibniz algebras and their associated gauge theories have already been discussed in 
the recent literature \cite{Hohm:2018ybo,Kotov:2018vcz}, notably in \cite{Hohm:2019wql}, where the notation of this paper 
is employed. However, the tensor hierarchies have typically 
only been developed up to the form degree needed in order to write a gauge invariant action. It would be important 
to use the general mathematical machinery defined here  to construct exact tensor hierarchies. 
In particular, this would allow one to formulate dynamical equations in terms of a hierarchy of duality relations between 
curvatures and their duals \cite{Bergshoeff:2009ph}. 

\item 
Apart from the applications in string and M-theory that motivated the formulation of Leibniz-Loday gauge theories,  
it is to be expected that they will play a role in other areas, too. For instance, it has recently been shown that these structures are needed 
for a local formulation of gauge theories based on the algebra of volume-preserving diffeomorphisms \cite{Hohm:2018git}, 
in turn suggesting potential applications in hydrodynamics \cite{ArnoldHydro}. 

\item Another area where applications are quite likely is that of higher-spin gauge theories as introduced  by Vasiliev, 
whose formulation relies on the unfolded approach, which is closely related to $L_{\infty}$ algebras \cite{Vasiliev:2005zu,Boulanger:2008up,Sharapov:2019vyd}, although interactions are more directly understood as governed by deformations of associative algebras \cite{Sharapov:2019vyd,Vasiliev:1999ba}.
One may suspect that the even further generalized algebraic structures discussed here will be useful for higher-spin gravity, particularly  in reference to the issue of modding out ideals that is crucial for higher-spin theories in arbitrary dimensions \cite{Vasiliev:2003ev,Arias:2017bvi,Bekaert:2017bpy,Grigoriev:2018wrx}.

\item We believe to have identified a new and rich mathematical  structure, but 
the formulation found here leaves something to be desired. For instance, it would be useful to understand the infinity enhanced Leibniz algebras 
as the `homotopy version' of some simpler algebraic structure --- in the same sense that an $L_{\infty}$ algebra is the homotopy 
version of a Lie algebra. Moreover, the arguably most efficient and useful formulation of $A_{\infty}$ or $L_{\infty}$ algebras 
is in terms of co-derivations on suitable tensor algebras that square to zero \cite{Lada:1992wc}. It would be helpful to find a similar 
formulation for the structures identified here.

\item 
An open problem in exceptional field theory, the duality covariant formulation of the spacetime actions of string/M-theory, 
is the question of whether there is a `universal' formulation unifying all U-duality groups, combining 
$E_{n(n)}$, $n=2,\ldots, 9$, into a single algebraic structure. So far, these theories are based on a split into `external' and `internal'  
spaces, with the latter governed by the Leibniz algebra of generalized diffeomorphisms and the former by $p$-forms building a tensor hierarchy 
for this  Leibniz algebra. Is there, perhaps, a formulation without split, based on a larger algebraic structure 
 from which one would recover the presently understood 
exceptional field theories by choosing a Leibniz subalgebra and decomposing according to a $\mathbb{Z}\oplus \mathbb{Z}$ 
grading as in (\ref{DIAGRAM})?

\end{itemize}

\subsection*{Acknowledgements}
We would like to thank Nicolas Boulanger, Martin Rocek, Henning Samtleben, Dennis Sullivan and Barton Zwiebach for useful discussions. We also like to thank an anonymous referee for a productive interaction that helped us to produce a significantly improved version of this manuscript.

This work is supported by the ERC Consolidator Grant ``Symmetries \& Cosmology".

\appendix

\section{Proof of the CS Bianchi identity}\label{App:CSproof}

In the construction of gauge covariant curvatures for the tensor hierarchy one is led to introduce Chern-Simons-like forms that are built from the one-form $A_1\equiv A$ alone, and are instrumental to prove the Bianchi identities. We define the pseudo Chern-Simons (CS) $n$-form by
\begin{equation}
\Omega_n(A)=\tfrac{(-1)^n}{(n-1)!}\,(\iota_A)^{n-2}\big[dA-\tfrac{1}{n}\,A\circ A\big]       \;,\quad |\Omega_n|=n-2\;,
\end{equation}
where
$\iota_A x:= A\bullet x$ and we recall that the pure Yang-Mills two-form $F_2$ is included as $\Omega_2\,$.
We will now prove the identity \eqref{CSrelations}
\begin{equation}\label{CSrelationsApp}
D\Omega_{n}+\tfrac12\sum_{k=2}^{n-1}\Omega_{k}\bullet\Omega_{n+1-k}=\cD\Omega_{n+1}\;,\quad n\geq 2\;,    
\end{equation}
that was used in section \ref{Sec:TensorH} to prove the Bianchi identity for the curvatures.
We will prove \eqref{CSrelationsApp} by induction. To this end, let us define the quantities
\begin{equation}
\begin{split}
\omega_n&:=\tfrac{(-1)^n}{(n-1)!}\, A\bullet(A\bullet(...(A\bullet dA)))\sim A^{n-2}dA  \;,\quad \omega_2\equiv dA\;,\\ 
a_n&:=\tfrac{(-1)^{n+1}}{n!}\, A\bullet(A\bullet(...(A\circ A)))\sim A^{n}\;,\quad a_2\equiv -\tfrac12\,A\circ A\;,
\end{split}
\end{equation}
 The pseudo CS form can then be written as
\begin{equation}\label{CSsmart}
\Omega_{n}=\omega_n+a_{n}\;, 
\end{equation}
and one has
\begin{equation}\label{DOmega}
D\Omega_{n}=d\omega_n+\Big[da_n-\cL_A\omega_n\Big]-\cL_A a_n\;    , 
\end{equation}
where we grouped terms with two, one and zero spacetime derivatives.
Notice that both $\omega_n$ and $a_n$ have form degree $n$ and intrinsic degree $n-2\,$, making them $\bullet$-commutative, \emph{i.e.}
\begin{equation}
\omega_k\bullet\omega_l=\omega_l\bullet\omega_k\;,\quad \omega_k\bullet a_l=a_l\bullet\omega_k\;,\quad a_k\bullet a_l=a_l\bullet a_k  \;,\quad \forall\;k,l\geq2  \;.
\end{equation}
From assumption 6) of \eqref{axioms}, upon combining internal and form degrees, one can derive
\begin{equation}
-\iota_A(\omega_k\bullet\omega_l)=(\iota_A\omega_k)\bullet\omega_l+\omega_k\bullet(\iota_A\omega_l)\;,     
\end{equation}
that holds also for $(a_k,a_l)$ and the mixed case $(\omega_k,a_l)\,$.
Finally,
by definition, they obey the recursive relation
\begin{equation}
\omega_{n+1}=-\tfrac{1}{n}\, A\bullet\omega_n\;,\quad a_{n+1}=-\tfrac{1}{n+1}\, A\bullet a_n\;,\quad \forall\; n\geq 2    \;,
\end{equation}
on which the proof is based.
We first show that
\begin{equation}\label{CSproof1}
d\omega_n+\tfrac12\sum_{k=2}^{n-1}\omega_k\bullet\omega_{n+1-k}=0\;,\quad\forall \;n\geq 2\;. \end{equation}
One has
\begin{equation}
\begin{split}
n=2 &\quad d\omega_2=d^2 A=0 \quad (\text{degenerate case})\\
n=3 &\quad d\omega_3=-\tfrac12\,d(A\bullet dA)=-\tfrac12\,dA\bullet dA=-\tfrac12\,\omega_2\bullet\omega_2\;.
\end{split}    
\end{equation}
Supposing that \eqref{CSproof1} holds for $n$ we can derive
\begin{equation}
\begin{split}
d\omega_{n+1}&=-\tfrac1n\,d(A\bullet\omega_n)=-\tfrac1n\,\omega_2\bullet\omega_n-\tfrac{1}{2n}\,A\bullet\Big[\sum_{k=2}^{n-1}\omega_k\bullet\omega_{n+1-k}\Big] \\
&= -\tfrac1n\,\omega_2\bullet\omega_n-\tfrac{1}{2n}\sum_{k=2}^{n-1}\Big[k\,\omega_{k+1}\bullet\omega_{n+1-k}+(n+1-k)\,\omega_k\bullet\omega_{n+2-k}\Big]\\
&= -\tfrac12\sum_{i=2}^n\omega_i\bullet\omega_{n+2-i}\;,
\end{split}    
\end{equation}
thus proving \eqref{CSproof1}.
By using it in \eqref{DOmega} we can write
\begin{equation}\label{DOmega'}
D\Omega_n+\tfrac12\sum_{k=2}^{n-1}\Omega_k\bullet\Omega_{n+1-k}=da_n-\cL_A\omega_n+\sum_{k=2}^{n-1}\omega_k\bullet a_{n+1-k}-\cL_Aa_n+\tfrac12\sum_{k=2}^{n-1}a_k\bullet a_{n+1-k}\;, 
\end{equation}
where we repeatedly used \eqref{CSsmart}. We now claim that
\begin{equation}\label{CSproof2}
da_n-\cL_A\omega_n+\sum_{k=2}^{n-1}\omega_k\bullet a_{n+1-k}=\cD\omega_{n+1}\;. \end{equation}
For the lowest values of $n$ one computes
\begin{equation}
\begin{split}
n=2\quad &da_2-\cL_A\omega_2=-\tfrac12\,d(A\circ A)-A\circ dA=-\tfrac12\,(A\circ dA+dA\circ A)\\
&=-\tfrac12\,\cD(A\bullet dA)=\cD\omega_3\quad\text{degenerate case}\;,\\
n=3\quad & da_3-\cL_A\omega_3+\omega_2\bullet a_2=\tfrac16\,d(A\bullet(A\circ A))+\tfrac12\,\cL_A(A\bullet dA)-\tfrac12\,(A\circ A)\bullet dA\\
&=-\tfrac13\,(A\circ A)\bullet dA-\tfrac16\,A\bullet(dA\circ A-A\circ dA)+\tfrac12\,\cL_A(A\bullet dA)\\
&=-\tfrac13\,(\cL_AA)\bullet dA+\tfrac13\,A\bullet(\cL_AdA)-\tfrac16\,A\bullet\cD(A\bullet dA)+\tfrac12\,\cL_A(A\bullet dA)\\
&=\tfrac16\,\cD(A\bullet(A\bullet dA))=\cD\omega_4\;. 
\end{split}    
\end{equation}
Supposing that \eqref{CSproof2} is valid for $n\,$, we deduce
\begin{equation}
\begin{split}
&da_{n+1}-\cL_A\omega_{n+1}+\sum_{k=2}^n\omega_k\bullet a_{n+2-k}=-\tfrac{1}{n+1}\,d(A\bullet a_n)-\cL_A\omega_{n+1}+\sum_{k=2}^n\omega_k\bullet a_{n+2-k}\\
&=-\tfrac{1}{n+1}\,\omega_2\bullet a_n+\tfrac{1}{n+1}\,A\bullet\Big[\cL_A\omega_n+\cD\omega_{n+1}-\sum_{k=2}^{n-1}\omega_k\bullet a_{n+1-k}\Big]-\cL_A\omega_{n+1}+\sum_{k=2}^n\omega_k\bullet a_{n+2-k}\\
&=-\tfrac{1}{n+1}\,\omega_2\bullet a_n-\tfrac{1}{n+1}\,[\cL_A(A\bullet\omega_n)+2\,a_2\bullet\omega_n]+\tfrac{1}{n+1}\,[\cL_A\omega_{n+1}-\cD(A\bullet\omega_{n+1})]-\cL_A\omega_{n+1}\\
&-\tfrac{1}{n+1}\sum_{k=2}^{n-1}[k\,\omega_{k+1}\bullet a_{n+1-k}+(n+2-k)\,\omega_k\bullet a_{n+2-k}]+\sum_{k=2}^n\omega_k\bullet a_{n+2-k}\\
&=-\tfrac{1}{n+1}\cD(A\bullet\omega_{n+1})=\cD\omega_{n+2}\;, 
\end{split}    
\end{equation}
thus proving the relation by induction.
The last step is to prove
\begin{equation}\label{CSproof3}
\cL_Aa_n-\tfrac12\sum_{k=2}^{n-1}a_k\bullet a_{n+1-k}=-\cD a_{n+1}\;.   
\end{equation}
The lowest values of $n$ give\footnote{Recall that, by using the Leibniz property, one has $(A\circ A)\circ A=2\,A\circ(A\circ A)\,$.} ($n=2$ is degenerate as usual)
\begin{equation}
\begin{split}
n=2\quad &\cL_A a_2=-\tfrac12\,\cL_A(A\circ A)=-\tfrac12\,A\circ(A\circ A)=-\tfrac16\,[A\circ (A\circ A)+(A\circ A)\circ A] \\
&=-\tfrac16\,\cD(A\bullet(A\circ A))=-\cD a_3\\
n=3\quad &\cL_Aa_3-\tfrac12\,a_2\bullet a_2=-\tfrac13\,\cL_A(A\bullet a_2)-\tfrac12\,a_2\bullet a_2=\tfrac16\,a_2\bullet a_2-\tfrac13\,A\bullet\cD a_3\\
&=-\tfrac13\,[\cL_Aa_3-\tfrac12\,a_2\bullet a_2]+\tfrac13\,\cD(A\bullet a_3)=\tfrac14\,\cD(A\bullet a_3)=-\cD a_4\;. 
\end{split}    
\end{equation}
Supposing that \eqref{CSproof3} holds for $n$ one has
\begin{equation}
\begin{split}
&\cL_Aa_{n+1}-\tfrac12\sum_{k=2}^na_k\bullet a_{n+2-k}=-\tfrac{1}{n+1}\,\cL_A(A\bullet a_n)-\tfrac12\sum_{k=2}^na_k\bullet a_{n+2-k}\\
&=\tfrac{2}{n+1}\,a_2\bullet a_n+\tfrac{1}{n+1}\,A\bullet\Big[\tfrac12\sum_{k=2}^{n-1}a_k\bullet a_{n+1-k}-\cD a_{n+1}\Big]-\tfrac12\sum_{k=2}^na_k\bullet a_{n+2-k}\\
&= \tfrac{2}{n+1}\,a_2\bullet a_n+\tfrac{1}{2(n+1)}\sum_{k=2}^{n-1}[(k+1)\,a_{k+1}\bullet a_{n+1-k}+(n+2-k)\,a_k\bullet a_{n+2-k}]\\
&-\tfrac{1}{n+1}\,[\cL_Aa_{n+1}-\cD(A\bullet a_{n+1})]-\tfrac12\sum_{k=2}^na_k\bullet a_{n+2-k}\\
&=-\tfrac{1}{n+1}\,\Big[\cL_Aa_{n+1}-\tfrac12\sum_{k=2}^na_k\bullet a_{n+2-k}\Big]+\tfrac{1}{n+1}\,\cD(A\bullet a_{n+1})=\tfrac{1}{n+2}\,\cD(A\bullet a_{n+1})=-\cD a_{n+2}\;, 
\end{split}    
\end{equation}
proving \eqref{CSproof3}. Using now the two results (\ref{CSproof2}), (\ref{CSproof3}) in \eqref{DOmega'} establishes the relation \eqref{CSrelations}.

\section{$L_\infty$ algebra from infinity enhanced Leibniz algebra}

Having discussed topological higher gauge theories based on a general $L_\infty$ algebra, we will show here how to construct a family of $L_\infty$ algebras from the data $(\circ, \cD, \bullet)$ of an infinity enhanced Leibniz algebra. Rather than discussing the most general $L_\infty$ algebra that can be constructed this way, we will make the choices that yield the simplest form for the $l_n$ brackets, and present them explicitly, acting on elements of arbitrary degree in the graded vector space $X\,$, up to the four-bracket $l_4\,$. 
\\
From now on we are going to distinguish elements $x,y,z,...$ in the $X_0$ subspace (that is the only one endowed with the Leibniz product $\circ$) from elements of higher degrees in $\bar X=\oplus_{n=1}^\infty X_n\,$, that will be denoted as $u_n\,$, with degree $|u_n|=n>0
\,$. The degree $-1$ nilpotent operator $l_1$ of the $L_\infty$ algebra  will be identified throughout this section with the $\cD$ operator of the Leibniz algebra, $l_1:=\cD$, that does not act on the subspace $X_0\,$.

\paragraph{$l_2$ brackets and $N=2$ relations}

We start from the $l_2$ bracket acting on two degree zero elements $x$ and $y\,$, that is completely fixed, up to an overall normalization, by degree and antisymmetry:
\begin{equation}\label{l2xy}
l_2(x,y):=[x,y]\equiv\tfrac12\,(x\circ y-y\circ x)\;.    
\end{equation}
Since $\cD$ does not act on $X_0\,$, there is no nontrivial $N=2$ relation at this level. The most general ansatz for the remaining $l_2$ brackets is given by
\begin{equation}\label{l2ansatz}
\begin{split}
&l_2(x,u_n):=k_n\,\cL_xu_n+j_n\,\cD(x\bullet u_n)\;,\quad l_2(u_n,x):=-l_2(x,u_n)\;,\quad n>0\;,\\
&l_2(u_n,u_m):=A_{nm}\,(u_n\bullet\cD u_m-(-1)^{nm}u_m\bullet\cD u_n)\;,\quad A_{nm}=A_{mn}\;,\quad n,m>0\;,
\end{split}    
\end{equation}
and is determined by degree and graded antisymmetry. The above brackets have to obey the $N=2$ relation $l_1l_2=l_2l_1\,$, that takes the explicit form
\begin{equation}\label{N=2relations}
\cD l_2(x,u_n)=l_2(x,\cD u_n)\;,\quad \cD l_2(u_n,u_m)=l_2(\cD u_n,u_m)+(-1)^nl_2(u_n,\cD u_m)\;.    
\end{equation}
By using covariance of the Lie derivative and nilpotency of $\cD$ one finds
\begin{equation}
\begin{split}
\cD l_2(x,u_n)&= k_n\,\cL_x\cD u_n \;,\\
l_2(x,\cD u_n)&=\left\{\begin{array}{l}
    \tfrac12\,\cL_x \cD u_n\;,\quad n=1\\
    k_{n-1}\,\cL_x\cD u_n+j_{n-1}\,\cD(x\bullet\cD u_n)=(k_{n-1}+j_{n-1})\cL_x\cD u_n\;,\quad n>1\;,
\end{array}\right.
\end{split}    
\end{equation}
from which one concludes that 
\begin{equation}
k_1=\tfrac12\;,\quad k_n=k_{n-1}+j_{n-1}=...=\tfrac12+\sum_{k=1}^{n-1}j_k\;,\quad n>1\;,    
\end{equation}
with all the $j_k$ parameters left free. We see that at each stage one has to single out the special cases of one (or multiple) $u_n$ elements having degree $+1\,$, since the corresponding $\cD u_n$ terms have degree zero, and thus have brackets of a different form. The same happens in order to verify the second relation in \eqref{N=2relations}:
\begin{equation}
\begin{split}
& \cD l_2(u_n,u_m)= -[1+(-1)^{n+m}]\,A_{nm}\, \cD u_n\bullet
\cD u_m\;,\\
& l_2(\cD u_n,u_m)-(-1)^{nm}l_2(\cD u_m,u_n)=\left\{\begin{array}{ll}
-2j_1\,\cD u_n\bullet\cD u_m\;,& n=m=1\\
-[j_m+(-1)^mA_{1\,m-1}]\,\cD u_n\bullet\cD u_m\;,& n=1\;, m>1\\
\big[A_{n-1\,m}+(-1)^{n+m}A_{n\,m-1}\big]\cD u_n\bullet\cD u_m\;,& n,m >1
\end{array}\right.
\end{split}    
\end{equation}
where we used $\cL_{\cD u}=0$ and the twisted Leibniz property 5) of \eqref{axioms}. The above result enforces $A_{11}=j_1$ and
\begin{equation}
\begin{split}
&[1-(-1)^m]\,A_{1m}=j_m+(-1)^mA_{1\,m-1}\;,\quad m>1\;,\\
&[1+(-1)^{n+m}]\,A_{nm}=-[A_{n-1\,m}+(-1)^{n+m}A_{n\,m-1}]\;,\quad n,m>1\;.
\end{split}    
\end{equation}
The first equation is solved by
\begin{equation}
A_{1n}=(-1)^nj_{n+1}+[1+(-1)^n]j_{n+2}\;,\quad n\geq1\;,\quad j_2=-j_1\;,    
\end{equation}
where $j_2=-j_1$ is required from matching $A_{11}=j_1$ with the first equation for $m=2\,$, and the general solution for $A_{1n}$ is found upon splitting the first equation above for the cases of even and odd values of $m\,$. Similarly, one can split the second equation into
\begin{equation}
\begin{split}
&A_{n\,m-1}=A_{n-1\,m}\;,\quad n+m \;{\rm odd\,,}\\
& A_{n\,m-1}+A_{n-1\,m}=-A_{nm}\;,\quad n+m \;{\rm even\,.}
\end{split}    
\end{equation}
The first case yields 
\begin{equation}
n+m \;{\rm even}\;\rightarrow\; A_{nm}=A_{n-1\,m+1}=...=A_{1\,n+m-1}=-j_{n+m} \end{equation}
that, used in the second equation, gives
\begin{equation}
n+m \;{\rm odd}\;\rightarrow\;    A_{nm}=-A_{n-1\,m+1}+2\,j_{n+m+1}\;.
\end{equation}
This imposes the simultaneous conditions
\begin{equation}
n+m \;{\rm odd}\;\rightarrow\; A_{nm}=j_{n+m}+2\,j_{n+m+1}=-j_{n+m}\;,     
\end{equation}
determining the most general solution for the $l_2$ brackets:
\begin{equation}\label{generall2}
\begin{split}
& j_{2k}=-j_{2k-1}\;,\quad k\geq1\;,\quad  j_{2k-1}\quad\text{are free parameters} \\
&k_n=\tfrac12+\tfrac{1+(-1)^n}{2}\,j_{n-1}\;,\quad n\geq1\;,\\
& A_{nm}=-j_{n+m}\;,\quad n,m\geq1\;.
\end{split}    
\end{equation}
Rather than using the general solution \eqref{generall2}, in the following we will fix the free parameters $j_n=0$ for all $n\geq1$ yielding the simplest set of $l_2$ brackets:
\begin{equation}\label{l2s}
\begin{split}
&l_2(x,y)=[x,y]=\tfrac12(\cL_xy-\cL_yx)\;,\quad x,y\in X_0\;,\\ 
&l_2(x,u_n)=\tfrac12\,\cL_xu_n\;,\quad n>0\;,\\
&l_2(u_n,u_m)=0\;,\quad n,m>0\;.
\end{split}    
\end{equation}
Before moving to the $l_3$ brackets, we use \eqref{l2s} to compute the Jacobiators, \emph{i.e.} the failure of the graded Jacobi identity:
\begin{equation}\label{jacobiators}
\begin{split}
&{\rm Jac}(x_1,x_2,x_3):=3\,l_2(l_2(x_{[1},x_2),x_{3]})=\tfrac12\,\cD[x_{[1}\bullet(x_2\circ x_{3]})] \;,\\   
&{\rm Jac}(x_1,x_2,u_n):=l_2(l_2(x_1,x_2),u_n)+2\,l_2(l_2(u_n,x_{[1}),x_{2]})=\tfrac14\,\cL_{[x_1,x_2]}u_n\;,\quad n>0\;,\\
&{\rm Jac}(x,u_n,u_m):=l_2(l_2(u_n,u_m),x)+2\,l_2(l_2(x,u_{[n}),u_{m)})=0\;,\quad n,m>0\;,\\
&{\rm Jac}(u_n,u_m,u_l):=3\,l_2(l_2(u_{[n},u_m),u_{l)})=0\;.
\end{split}    
\end{equation}
To derive the first relation we have used the identity
\begin{equation}\label{nestedcircles}
x_{[1}\circ(x_2\circ x_{3]})=\tfrac12\,(x_{[1}\circ x_2)\circ x_{3]}=2\,[[x_{[1},x_2],x_{3]}]=\tfrac13\,\cD(x_{[1}\bullet(x_2\circ x_{3]}))\;,
\end{equation}
that originates from the Leibniz property, and we have used the shorthand notation $[nm)$ and $[nml)$ for total graded antisymmetrization with strength one, \emph{i.e.} $T_{[nm)}:=\tfrac12\,(T_{nm}-(-1)^{nm}T_{mn})$ and so on.

\paragraph{$l_3$ brackets and $N=3$ relations}

The first Jacobiator in \eqref{jacobiators} uniquely fixes the three-bracket on three degree zero elements $x_1,x_2$ and $x_3$ from the $N=3$ relation
\begin{equation}
\cD l_3(x_1,x_2,x_3)+{\rm Jac}(x_1,x_2,x_3)=0\;,
\end{equation}
to be
\begin{equation}
l_3(x_1,x_2,x_3)=-\tfrac12\,x_{[1}\bullet(x_2\circ x_{3]})  \;.  
\end{equation}
For elements of higher degree, graded antisymmetry and the properties 5) and 6) of \eqref{axioms}, together with $|l_3|=+1$ fix the most general ansatz to be
\begin{equation}\label{l3ansatz}
\begin{split}
&l_3(x_1,x_2,u_n):=\alpha_n\,[x_1,x_2]\bullet u_n+\beta_n\,x_{[1}\bullet\cL_{x_{2]}}u_n+\gamma_n\,\cD(x_{[1}\bullet(x_{2]}\bullet u_n))\;,\quad n>0\;,\\
&l_3(x,u_n,u_m):=a_{nm}\,u_{[n}\bullet\cL_x u_{m)}+b_{nm}\,u_{[n}\bullet(x\bullet\cD u_{m)})+c_{nm}\,\cD(u_{[n}\bullet(u_{m)}\bullet x))\;,\quad n,m>0\;,\\
&l_3(u_k,u_n,u_m):=A_{knm}\,u_{[k}\bullet(u_n\bullet\cD u_{m)})\;,\quad k,n,m>0\;.
\end{split}    
\end{equation}
The relevant $N=3$ relations to be satisfied read
\begin{equation}\label{N=3relations}
\begin{split}
&\cD l_3(x_1,x_2,u_n)+l_3(x_1,x_2,\cD u_n)+\tfrac14\,\cL_{[x_1,x_2]}u_n=0\;,\quad n>0\;,\\ 
&\cD l_3(x,u_n,u_m)+2\,l_3(x,\cD u_{[n},u_{m)})=0\;,\quad n,m>0\;,\\   
&\cD l_3(u_k,u_n,u_m)+3\,l_3(\cD u_{[k},u_n,u_{m)})=0\;,\quad k,n,m>0\;.
\end{split}    
\end{equation}
In order to compute the above expressions from the ansatz \eqref{l3ansatz}, one has to treat separately the cases of one or more $u$'s having degree $+1\,$, just as it has been shown explicitly for the $l_2$ bracket. A straightforward but tedious computation shows that the only free parameter left, after imposing \eqref{N=3relations}, is $\gamma_1\,$. Instead of showing the entire proof, we rather choose $\gamma_1=0\,$, that gives the simplest realization of the brackets, and show that this choice is indeed consistent with \eqref{N=3relations}. With $\gamma_1$ fixed to zero, the $l_3$ brackets read
\begin{equation}\label{l3s}
\begin{split}
&l_3(x_1,x_2,x_3)=-\tfrac12\,x_{[1}\bullet(x_2\circ x_{3]})  \;,\\  
&l_3(x_1,x_2,u_n)=-\tfrac16\,[x_1,x_2]\bullet u_n-\tfrac16\,x_{[1}\bullet\cL_{x_{2]}}u_n\;,\quad n>0\;,\\
&l_3(x,u_n,u_m)=\tfrac16\,u_{[n}\bullet\cL_x u_{m)}\;,\quad n,m>0\;,\\
&l_3(u_k,u_n,u_m)=0\;,\quad k,n,m>0\;.
\end{split}    
\end{equation}
For any $n>0$ one has
\begin{equation}
\cD l_3(x_1,x_2,u_n)=-\tfrac16\,\cD\big\{[x_1,x_2]\bullet u_n+x_{[1}\bullet\cL_{x_{2]}} u_n\big\}\;,    
\end{equation}
and
\begin{equation}
l_3(x_1,x_2,\cD u_n)= -\tfrac16\,[x_1,x_2]\bullet \cD u_n-\tfrac16\,x_{[1}\bullet\cL_{x_{2]}}\cD u_n\;, 
\end{equation}
where, for $n>1\,$, this is computed from $l_3(x_1,x_2,u_{n-1})$ with $u_{n-1}=\cD u_n\,$, while for $n=1$ we used $l_3(x_1,x_2,x_3)$ with $x_3=\cD u_1\,$. Summing the two contributions one obtains
\begin{equation}
\begin{split}
\cD l_3(x_1,x_2,u_n)+l_3(x_1,x_2,\cD u_n)&=-\tfrac16\,\cL_{[x_1,x_2]}u_n-\tfrac16\,\big[x_{[1}\bullet\cD(\cL_{x_{2]}}u_n)+\cD(x_{[1}\bullet\cL_{x_{2]}}u_n)\big]\\
&=-\tfrac16\,\cL_{[x_1,x_2]}u_n-\tfrac16\,\cL_{x_{[1}}\cL_{x_{2]}}u_n=-\tfrac14\,\cL_{[x_1,x_2]}u_n\;,
\end{split}
\end{equation}
thus proving the first of the relations \eqref{N=3relations}. Similarly, for any $n,m>0$ one has
\begin{equation}
\cD l_3(x,u_n,u_m)=\tfrac16\,\cD\big[u_{[n}\bullet\cL_x u_{m)}\big]\;,    
\end{equation}
while, for $n,m>1$ one has
\begin{equation}
\begin{split}
2\,l_3(x,\cD u_{[n},u_{m)})&\stackrel{[nm)}{=}\tfrac16\big[\cD u_{n}\bullet\cL_x u_{m}-(-1)^{(n-1)m}u_m\bullet\cL_x\cD u_n\big] \\
&\stackrel{[nm)}{=}\tfrac16\big[\cD u_{n}\bullet\cL_x u_{m}+(-1)^nu_n\bullet\cL_x\cD u_m\big]\\
&\stackrel{[nm)}{=}-\tfrac16\,\cD\big[u_{n}\bullet\cL_x u_{m}\big]
\end{split}    
\end{equation}
thus proving the relation for $n,m>1\,$. When $n=1$ and $m>1$ one obtains the same result:
\begin{equation}
\begin{split}
&l_3(x,\cD u_1,u_m)-l_3(x,u_1,\cD u_m)= -\tfrac{1}{12}\,\cL_x\cD u_1\bullet u_m+\tfrac{1}{12}\,\cD u_1\bullet\cL_x u_m\\
&\hspace{56mm}-\tfrac{1}{12}\,\big[u_1\bullet\cL_x \cD u_m+(-1)^m\cD u_m\bullet\cL_x u_1\big]  \\
&= \tfrac{1}{12}\,\big[\cD u_1\bullet\cL_x u_m-u_1\bullet\cL_x \cD u_m\big]-(-1)^m\tfrac{1}{12}\,\big[\cD u_m\bullet\cL_x u_1+(-1)^mu_m\bullet\cL_x\cD u_1\big]\\
&=-\tfrac{1}{12}\,\cD\big[u_1\bullet\cL_x u_m-(-1)^mu_m\bullet\cL_xu_1\big]\;,
\end{split}    
\end{equation}
with the first term computed from $l_3(x_1,x_2,u_m)$ for $x_2=\cD u_1\,$. Similarly, the same is also obtained for $n=m=1\,$:
\begin{equation}
\begin{split}
2\,l_3(x,\cD u_{(1},u_{2)})&=-\tfrac{1}{6}\,\cL_x\cD u_{(1}\bullet u_{2)}+\tfrac{1}{6}\,\cD u_{(1}\bullet\cL_x u_{2)} \\
&=\tfrac16\big[\cD u_{(1}\bullet\cL_x u_{2)}-u_{(1}\bullet\cL_x\cD u_{2)}\big]\\
&=-\tfrac16\,\cD\,\big[u_{(1}\bullet\cL_x u_{2)}\big]\;,
\end{split}    
\end{equation}
where both $|u_1|=|u_2|=1$ and we used the subscripts to distinguish the elements, rather than denoting the degree. Finally, the last relation in \eqref{N=3relations} does not need any computation to be proved, since either every term is identically zero or, if any element of degree one is present, the corresponding degree zero object $\cD u$ only acts through a Lie derivative, and thus vanishes. With this we have thus proved that \eqref{l3s} provides a consistent set of three-brackets on the entire space.

\paragraph{$l_4$ brackets and $N=4$ relations}

As the last explicit realization of the $l_n$ brackets, we will now show that, given the two- and three-brackets as in \eqref{l2s} and \eqref{l3s}, \emph{all} the four brackets $l_4$ vanish. The abstract $N=4$ relation reads $l_1l_4-l_4l_1=l_2l_3-l_3l_2$ so, rather than giving a general ansatz for the $l_4$ maps, we will prove that $l_2l_3=l_3l_2$ on the entire space. From the initial relation
\begin{equation}
\cD l_4(x_1,x_2,x_3,x_4)=4\,l_2(l_3(x_{[1},x_2,x_3),x_{4]})-6\,l_3(l_2(x_{[1},x_2),x_3,x_{4]})=0    
\end{equation}
one can then prove recursively that all $l_4$ maps vanish.
In order to prove $l_2l_3=l_3l_2\,$, we start indeed with all four elements in $X_0\,$, giving
\begin{equation}
\begin{split}
&4\,l_2(l_3(x_{[1},x_2,x_3),x_{4]})-6\,l_3(l_2(x_{[1},x_2),x_3,x_{4]})\\
\stackrel{[1234]}{=}&\cL_{x_4}\big[x_1\bullet\cL_{x_2}x_3\big]+[x_1,x_2]\bullet[x_3,x_4]+2\,x_3\bullet[x_4,[x_1,x_2]]    \\
\stackrel{[1234]}{=}&-\cL_{x_1}\big[x_2\bullet\cL_{x_3}x_4\big]+(\cL_{x_1}x_2)\bullet(\cL_{x_3}x_4)+x_2\bullet[x_1\circ(x_3\circ x_4)]=0\;,
\end{split}    
\end{equation}
that proves $l_4(x_1,x_2,x_3,x_4)=0\,$. The next quadratic relation reads (antisymmetrization $[123]$ is understood)
\begin{equation}
\begin{split}
&l_2(l_3(x_1,x_2,x_3),u_n)-3\,l_2(l_3(x_1,x_2,u_n),x_3)-3\,l_3(l_2(x_1,x_2),x_3,u_n)-3\,l_3(l_2(x_1,u_n),x_2,x_3)\\
&= \tfrac32\,\cL_{x_3}l_3(x_1,x_2,u_n)+\tfrac14\big\{[(x_1\circ x_2)\circ x_3-x_3\circ(x_1\circ x_2)]\bullet u_n+(x_1\circ x_2)\bullet\cL_{x_3}u_n\\
&-x_3\bullet\cL_{[x_1,x_2]}u_n+(x_2\circ x_3)\bullet\cL_{x_1}u_n+x_2\bullet\cL_{x_3}\cL_{x_1}u_n\big\}\\
&= \tfrac32\,\cL_{x_3}l_3(x_1,x_2,u_n)+\tfrac14\big\{[x_3\circ(x_1\circ x_2)]\bullet u_n+(x_1\circ x_2)\bullet\cL_{x_3}u_n\\
&+(x_3\circ x_1)\bullet\cL_{x_2}u_n+x_1\bullet\cL_{x_3}\cL_{x_2}u_n\big\}\\
&= \tfrac32\,\cL_{x_3}l_3(x_1,x_2,u_n)+\tfrac14\,\cL_{x_3}\big\{(x_1\circ x_2)\bullet u_n+x_1\bullet\cL_{x_2}u_n\big\}=0\;,
\end{split}    
\end{equation}
proving $l_4(x_1,x_2,x_3,u_n)=0\,$. Next, when two elements have degree higher than zero one has
\begin{equation}
\begin{split}
&2\,l_2(l_3(x_1,x_2,u_n),u_m)+2\,l_2(l_3(x_1,u_n,u_m),x_2)\\
&-l_3(l_2(x_1,x_2),u_n,u_m)+4\,l_3(l_2(x_1,u_n),x_2,u_m)-l_3(l_2(u_n,u_m),x_1,x_2) \\
&=\cL_{x_1}l_3(x_2,u_n,u_m)-\tfrac16\,u_n\bullet\cL_{[x_1,x_2]}u_m+2\,l_3(x_1,\cL_{x_2}u_n,u_m)\\
&=\cL_{x_1}l_3(x_2,u_n,u_m)-\tfrac16\,u_n\bullet\cL_{[x_1,x_2]}u_m+\tfrac16\,\cL_{x_2}u_n\bullet\cL_{x_1}u_m-(-1)^{mn}\tfrac16\,u_m\bullet\cL_{x_1}\cL_{x_2}u_n\\
&=\cL_{x_1}l_3(x_2,u_n,u_m)-\tfrac16\,\cL_{x_1}\big[u_n\bullet\cL_{x_2}u_m\big]=0
\end{split}    
\end{equation}
with $[12]$ and $[nm)$ left implicit, yielding $l_4(x_1,x_2,u_n,u_m)=0\,$. The last two cases, namely $l_4(x,u_k,u_n,u_m)$ and $l_4(u_k,u_l,u_m,u_n)$ do not need any computation, since any term in $l_2l_3$ and $l_3l_2$ vanishes identically. This finally proves that, with the choice \eqref{l2s} and \eqref{l3s} for the lower brackets, all the $l_4$'s vanish. Moreover, it can be shown that the $l_4$ brackets vanish for any choice of the $l_3$'s, provided that the $l_2$ maps are given by \eqref{l2s}.

We summarize here the list of non-vanishing $L_\infty$ brackets explicitly constructed from the Leibniz algebra thus far:\footnote{\emph{Note added in proof}: In the meantime, it was shown in \cite{Lavau:2019oja} how to construct the associated $L_\infty$ brackets in general by using the derived bracket construction.}
\begin{equation}\label{allls}
\begin{split}
&l_2(x,y)=[x,y]\;,\quad x,y\in X_0\;,\\ 
&l_2(x,u_n)=\tfrac12\,\cL_xu_n\;,\quad n>0\;,\\[1mm]
&l_3(x_1,x_2,x_3)=-\tfrac12\,x_{[1}\bullet(x_2\circ x_{3]})  \;,\\ &l_3(x_1,x_2,u_n)=-\tfrac16\,[x_1,x_2]\bullet u_n-\tfrac16\,x_{[1}\bullet\cL_{x_{2]}}u_n\;,\quad n>0\;,\\
&l_3(x,u_n,u_m)=\tfrac16\,u_{[n}\bullet\cL_x u_{m)}\;,\quad n,m>0\;.
\end{split}    
\end{equation}
Having shown that all $l_4$ brackets vanish does not mean that higher brackets vanish, and indeed one can prove that there is no allowed choice 
of coefficients for which all $l_5$ maps are zero. 
In particular, the richer structure on the Leibniz side hints at the existence of special points in the ``moduli space'' of $L_\infty$ maps, for which infinitely many brackets vanish, even though the $L_\infty$ algebra is not truncated to a finite degree.

\end{document}